\DocumentMetadata{
	lang		= en-US,
	pdfversion  = 1.7,
	pdfstandard = a-2b,
}
\PassOptionsToPackage{table}{xcolor}
\documentclass[twoside]{mitthesis}

\usepackage{listings}
\usepackage{upquote}
\usepackage[version=4]{mhchem}
\usepackage{lipsum}
\IfPackageAtLeastTF{lipsum}{2021/09/20}{\setlipsum{auto-lang=false}}{}
\usepackage{booktabs}
\usepackage{array}
\usepackage{dcolumn}
  \newcolumntype{d}[1]{D{.}{.}{#1}}
\usepackage{longtable}
\usepackage[nopatch=footnote]{microtype}
\usepackage{multicol}

\usepackage[style=ext-numeric-comp,giveninits=true,maxbibnames=10,sorting=none,language=american]{biblatex}

    \AtEveryBibitem{%
      \ifentrytype{article}{%
        \renewbibmacro{in:}{}%
		\renewbibmacro*{issue+date}{%
		\iffieldundef{issue}%
        	{}%
        	{\printfield{issue}%
         	\setunit*{\addcomma\space}
        	}%
			\printdate
		}
      }{}
    }
	\renewbibmacro*{volume+number+eid}{%
        \printtext[bold]{\printfield{volume}}%
        \iffieldundef{number}
          {} %
          {\printtext[parens]{\printfield{number}}} %
        \setunit{\bibeidpunct}%
        \printfield{eid}
    }

\usepackage{setspace}

\hypersetup{%
	pdfsubject={An evaluation of LLM-driven vericoding in Lean and the design of agent-directed tree-search methods},
	pdfkeywords={Leo Yao, Massachusetts Institute of Technology, MIT, formal verification, verified-code generation, vericoding, automated proof synthesis, Lean theorem prover, mathlib, vericoding-benchmark, large language models, LLM agents, agentic loop, tree search, inference-time search},
	pdfcontactemail={leoy@mit.edu},
}

\DeclareMathOperator{\Erf}{erf}
\DeclareMathOperator{\Erfc}{erfc}
\DeclarePairedDelimiterXPP\erf[1]{\Erf\mkern1mu}(){}{#1}
\DeclarePairedDelimiterXPP\erfc[1]{\Erfc\mkern1mu}(){}{#1}

\begin{document}
\title{Automating Formal Verification with Agent-Guided Tree Search}
\Author{Leo Yao}{Department of Electrical Engineering and Computer Science}[S.B. in Physics and Electrical Engineering and Computer Science, Massachusetts Institute of Technology, 2025]
\Degree{Master of Engineering in Electrical Engineering and Computer Science}{Department of Electrical Engineering and Computer Science}
\Supervisor{Max Tegmark}{Professor of Physics}[Department of Physics]
\Acceptor{Katrina LaCurts}{Chair, Master of Engineering Thesis Committee}{}
\DegreeDate{May}{2026}
\ThesisDate{May 22, 2026}
\maketitle

\begin{abstract}
	Formal verification offers a path to provably correct software, but writing verified code remains expensive enough that the technique is rarely used in production. Recent large language models can accelerate this work, and recent benchmarks measure their ability to translate specifications into code and machine-checked proofs of correctness.
	This thesis evaluates the state of such LLM-driven verified-code generation (``vericoding'') in Lean and develops search-based methods for improving verification performance.
	We first reproduce a subset of the \texttt{vericoding-benchmark} Lean leaderboard on a current cross-vendor model pool, finding that non-reasoning performance remains roughly steady on US closed-source models while open-weight models have slightly improved. We update the iterative methodology of \texttt{vericoding-benchmark} with an agentic loop equipped with mathlib search, finding that model performance greatly improves and scales with agent budget. GPT-5.4 nearly saturates the benchmark at 95.0\% on 423 specs with $K=50$ LLM calls.
	We then design two agent-directed tree-search formulations: a state-based orchestrator that branches on partial-proof states, and a context-based orchestrator that branches on full subagent contexts.
	Compared against the agent baseline, the context-based design solves a wider range of intermediate-difficulty specs at lower token cost, while the agent baseline retains an advantage on the hardest specs, where uninterrupted iteration matters most.
	We conclude that search structure has selective advantages over a strong agent baseline, and that more challenging benchmarks drawn from modern code are important to measure and drive further progress in automated formal verification.

	Code available upon request by contacting the author at \texttt{leoy@mit.edu}.
\end{abstract}

\chapter*{Acknowledgments}
\pdfbookmark[0]{Acknowledgments}{acknowledgments}

I would like to thank my research advisor, Professor Max Tegmark, for four years of guidance and support.
Thank you for introducing me to numerous areas of study across physics, machine learning, and formal verification.
It has been an honor to be part of your group for the majority of my time at MIT.

Thank you to research group members David Baek, Max Tan, and Ionel-Emilian Chiosa for insightful discussions, contributions to research direction, and technical assistance; collaborators at the Beneficial AI Foundation for their mentorship and for building the benchmark this thesis stands on; and staff in the MIT Department of Physics and MIT EECS Department for their administrative assistance.

I am grateful to my undergraduate research supervisor, Dr. Ziming Liu, for multiple years of dedicated mentorship and helping shape me as a researcher.
I am also grateful to Professor Soonwon Choi for thoughtful guidance as my undergraduate academic advisor and Professor Michael Sipser for taking me on as a teaching assistant.

Finally, I would like to thank my family and friends for their continued dedication, understanding, and support.

\tableofcontents
\listoffigures
\listoftables

\doublespacing

\chapter{Introduction}

With the development and growing application of artificial intelligence (AI) in recent years, the question of safety has grown from a niche research area into a mainstream concern~\cite{Bengio_2024}.
A key technical question in AI safety is trustworthiness: whether AI systems are faithful to human goals, and reliably act in support of those goals when exposed to different inputs and environments.
This question becomes more important as models become more powerful and are deployed with increasing autonomy: unlike the purely chat-interface models of a few years ago, present-day models are run as agents which can write code, interact with the web, and execute actions on computer systems. AI agents have been able to complete longer and more complicated tasks, with a measure of completed task length doubling every seven months~\cite{10.48550/arxiv.2503.14499,metr_time_horizon_1_1_2026,metr_time_horizons_live}.
One possible tail-risk scenario is that of recursive self-improvement, in which improvements feed back into AI research itself. In such a regime, research progress and capabilities accelerate exponentially, possibly outpacing the techniques and safeguards used to keep up with and control them~\cite{10.48550/arxiv.2306.12001}.
Even today, techniques used to align models do not fully mitigate the risks of unintended behavior and misuse. Agents have been observed taking destructive actions within compute environments or inadvertently publicizing sensitive information, and current models can supply guidance on biological or chemical weapons or generate harmful content~\cite{anthropic_mythos_system_card_2026,openai_gpt55_system_card_2026}.

Coding in particular has been broadly transformed by recent developments in AI capabilities. Current generation large language models (LLMs) excel at generating large quantities of code from oftentimes vague natural language descriptions, a practice commonly referred to as vibe coding~\cite{10.48550/arxiv.2507.21928}.
While such code often looks correct at first glance, it may contain bugs or security vulnerabilities that are not immediately apparent~\cite{Pearce_2022,10.48550/arxiv.2604.05292}.
The conventional approach is to use collections of tests, from individual unit tests of functions to larger end-to-end tests of behavior, to attempt to check correctness. Modern coding agents are able to autonomously write and run tests to try to detect and correct errors.
However, AI-assisted software development is still marked by frequent hallucinations and deviations from the expected task, leading to behavior such as avoiding or changing the tests, hardcoding success cases, or even pretending that tests have passed when they have not.

An alternative is \emph{formal verification}: a method to prove the correctness of mathematical theorems or computer programs, written in formal languages that can be checked by computers.
These methods were first developed for mathematics, to allow for computer-assisted validation of long or intricate proofs.
The most active language ecosystem is Lean 4, along with the community-maintained library mathlib, which has formalized substantial portions of undergraduate- and graduate-level mathematics~\cite{Moura_2021}.
Human-formalized collections of questions drawn from competition mathematics, such as miniF2F~\cite{10.48550/arxiv.2109.00110}, PutnamBench~\cite{10.48550/arxiv.2407.11214}, and various IMO (International Mathematics Olympiad) formalizations, serve as benchmarks for automated methods.
AI capabilities have rapidly progressed in the last few years, with systems such as DeepMind's AlphaProof reaching human-level IMO performance~\cite{Hubert_2025}.

Compared to test-based approaches, formal verification has the advantage of guaranteeing that an implementation adheres to its full specification.
This guarantee matters more now that frontier models can accelerate the process of finding and exploiting vulnerabilities~\cite{anthropic_mythos_system_card_2026,openai_gpt55_system_card_2026}---the best way to avoid exploits is to prove them impossible.
Verified code is not a hypothetical: provably correct software ships today in domains from cryptography to kernels to compilers~\cite{Zinzindohoue_2017,Klein_2009,Leroy_2009}.
However, each of these projects took significant human expert effort, which is why formal verification remains rare in production systems.

Recent work on LLM-driven formal verification spans reinforcement learning~\cite{10.48550/arxiv.2202.01344}, subgoal decomposition~\cite{10.48550/arxiv.2509.22819}, and tree search over proof states~\cite{10.48550/arxiv.2205.11491,Hubert_2025}, alongside the tooling and library integrations that make these methods usable.
However, most research focuses on mathematical benchmarks, while verified coding is comparatively less studied.
This thesis studies the application of LLM-driven methods to formal verification problems in code, devising and testing various agent-driven search approaches to construct proofs.

The rest of the thesis proceeds as follows. In Chapter~\ref{ch:literature}, we review the literature surrounding current language models, formal verification, and machine learning methods for Lean. In Chapter~\ref{ch:baselines}, we provide updated baselines for Lean verification using both the original iterative prompting approach and an agent-based framework. In Chapter~\ref{ch:search}, we introduce two agent-guided search methods and evaluate them in comparison with our baselines. We conclude in Chapter~\ref{ch:conclusion} with directions for future work.

\chapter{Literature Review}
\label{ch:literature}

\section{Current Language Models}

Modern LLMs are built on the transformer architecture, which succeeds previous recurrent architectures with a self-attention mechanism that can be parallelized in training~\cite{10.48550/arxiv.1706.03762}.
Current frontier models are decoder-only autoregressive transformers~\cite{10.48550/arxiv.2005.14165}, which generate text one token at a time, conditioning each new token on all preceding context.
These models are first pretrained on massive corpora of unstructured text drawn from much of the public internet, in which the model learns to predict the next token from previous ones.
It has been found empirically that these models obey scaling laws: loss decreases as a power-law in compute, dataset size, and parameter count~\cite{10.48550/arxiv.2001.08361,10.48550/arxiv.2203.15556}. This regularity has driven rapid increases in model size and training compute in recent years.

Without further training, pretrained LLMs produce raw continuations of input rather than answering questions or following user instructions. To adapt these models into useful assistants, an additional post-training step is required.
The standard approach is reinforcement learning from human feedback (RLHF), in which the model is fine-tuned against a reward model trained on human preferences~\cite{10.48550/arxiv.1706.03741,10.48550/arxiv.2203.02155}. Variants such as Constitutional AI reduce reliance on human labels by combining model self-critique with AI-generated preference feedback~\cite{10.48550/arxiv.2212.08073}, and direct preference optimization (DPO) collapses RLHF's two-stage pipeline into a single preference-based classification loss~\cite{10.48550/arxiv.2305.18290}.

Additional improvements in capabilities come from inference-time reasoning, where models generate intermediate reasoning steps in a chain-of-thought before returning final responses~\cite{10.48550/arxiv.2201.11903}. This substantially improves performance on multi-step tasks, and also scales with tokens spent on thinking. Recent training methods reward final-answer correctness and let the model learn its reasoning process via reinforcement learning~\cite{10.48550/arxiv.2501.12948,openai_o1_system_card_2024}.

Beyond just responding to user inputs, recent LLMs are increasingly deployed as agents that can access and modify external systems. They do so through tool use, which exposes external functions to the agent through structured requests.
This capability was advanced by methods such as ReAct, which interleaves reasoning steps with tool calls~\cite{10.48550/arxiv.2210.03629}, and Toolformer, which trains models to invoke tools without supervision~\cite{10.48550/arxiv.2302.04761}. Common tools now include web search for retrieving up-to-date information, script and shell access for running code, and connections to third-party apps and APIs.
The standard interface for tool integration is the Model Context Protocol (MCP), originally introduced by Anthropic in late 2024~\cite{anthropic_mcp_announcement_2024} and since donated to the Linux Foundation's Agentic AI Foundation~\cite{anthropic_mcp_aaif_donation_2025}. The protocol allows developers to expose toolsets to any compatible client via dedicated MCP servers.
These capabilities are increasingly available to broader audiences through consumer-facing agents such as Claude Code~\cite{anthropic_claude_code}, OpenAI's Codex~\cite{openai_codex}, and the open-source OpenCode~\cite{opencode}, and are increasingly integrated into development environments and other end-user applications.

Between scaling and integration of external tools, model capabilities have advanced rapidly. On METR's time horizon benchmark, the frontier task length at 50\% completion reliability has roughly doubled every four to seven months, reaching around fourteen hours for current leading models~\cite{10.48550/arxiv.2503.14499,metr_time_horizon_1_1_2026,metr_time_horizons_live}.
Just this year, reported offensive cyber capabilities have risen sharply. Anthropic's Claude Mythos Preview demonstrated the ability to autonomously discover and exploit zero-day vulnerabilities in major operating systems and browsers, and was held back from public release~\cite{anthropic_mythos_system_card_2026}.
OpenAI's GPT-5.5, released in April 2026, also demonstrated a step change in offensive cyber capability, and is rated ``High'' for cybersecurity under the company's Preparedness Framework, prompting expanded safeguards under OpenAI's Trusted Access for Cyber program~\cite{openai_gpt55_system_card_2026}.
These capabilities pose increasing risks to existing software infrastructure, and the underlying rate of progress may itself continue to accelerate.

\section{Formal Verification}

Proof assistants were first developed for mathematics in the late 1980s and 1990s, to automate the verification of lengthy arguments or extensive case analyses. Each is built around a small logical kernel that mechanically checks every step, an architecture originating in Stanford LCF~\cite{milner_lcf_1972} and consolidated in Edinburgh LCF~\cite{Gordon_1979}, and reusable libraries of definitions and theorems.
Coq (recently renamed Rocq) is based on the Calculus of Inductive Constructions~\cite{Coquand_1990}, and has been used for landmark pure-math formalizations such as the Four-Color~\cite{gonthier_four_color_2008} and Feit-Thompson theorems~\cite{Gonthier_2013}, as well as verified software including the C compiler CompCert~\cite{Leroy_2009}.
Isabelle/HOL is built on classical higher-order logic~\cite{Nipkow_2002} and maintains the Archive of Formal Proofs, a large community-curated library of formalized results~\cite{archive_of_formal_proofs}; it is also the system used to verify the seL4 microkernel~\cite{Klein_2009}.
Agda treats proofs and programs uniformly, with a clean functional-language syntax and dependent pattern matching~\cite{norell_agda_thesis_2007}.

Lean was created by Leonardo de Moura at Microsoft Research in 2013~\cite{de_Moura_2015}, combining a small trusted dependent-type-theory kernel with support for both interactive and automated theorem proving. Lean 3 introduced a metaprogramming framework in which proof procedures are themselves Lean programs, in contrast to Coq's Ltac domain-specific tactic language~\cite{Ebner_2017}.
The community-maintained library mathlib, factored out of Lean's core library in 2017, consolidates definitions and theorems within a single repository, with shared dependencies and consistent style~\cite{The_mathlib_Community_2020}. By 2025 it contained over two million lines of formalized mathematics. It also supports notable standalone projects such as Scholze's Liquid Tensor Experiment~\cite{liquid_tensor_experiment} and the Polynomial Freiman-Ruzsa conjecture~\cite{pfr_project}.
Lean 4~\cite{Moura_2021}, released in 2021, rewrites the kernel, elaborator, and tactic framework in Lean itself, creating a unified language for both proofs and programs. The community port of mathlib to Lean 4 completed in 2023.
The Lean ecosystem consists of tooling that exposes proof state programmatically: tactic libraries such as \texttt{Aesop} provide configurable best-first proof search for Lean~\cite{Limperg_2023}, and the Lean language server exposes elaboration state, goals, and diagnostics over a uniform protocol. A more recent Model Context Protocol bridge wraps the language server, making this proof state directly queryable by LLM agents~\cite{lean_lsp_mcp}.
Beyond mathematics, Lean is increasingly being used for code verification, in projects such as the Rust-to-Lean \texttt{Aeneas} pipeline~\cite{Ho_2022,aeneas_lean_backend} and \texttt{VCV-io}, a framework for cryptographic-protocol proofs~\cite{vcv_io_iacr_2024}. This combination of a unified language, a deep mathematical library, and a programmatic tooling ecosystem has made Lean 4 the focal point of current AI-driven proof methods.

Other code-focused verifiers, such as Dafny, F*, and Verus, replace user-written proofs with SMT-backed automation: the programmer annotates the program with specifications and invariants, and the solver discharges the resulting obligations. Dafny~\cite{Leino_2010} is the canonical example, underlying academic systems-verification projects such as \texttt{IronFleet} for distributed systems~\cite{Hawblitzel_2015} and \texttt{VeriBetrKV} for storage~\cite{hance_veribetrkv_osdi_2020}.
F*~\cite{Swamy_2016} is similarly SMT-backed but further extends the type system, underlying active cryptographic-verification research including \texttt{HACL*}~\cite{Zinzindohoue_2017}, whose verified code is reused in libraries such as \texttt{libcrux}~\cite{libcrux}.
Verus~\cite{Lattuada_2023} extends Rust with verification annotations while preserving the language's core structure, and has been applied in projects such as \texttt{OwlC} for cryptographic protocols~\cite{owlc_iacr_2025} and \texttt{dalek-lite} for the \texttt{curve25519-dalek} elliptic-curve library~\cite{dalek_lite}.

\section{Machine Learning Methods for Lean}

The earliest applications of machine learning to Lean were interfaces that allowed ML systems to inspect proof state and test tactic steps, used both to train autonomous provers and to assist humans. This approach originated with OpenAI's GPT-f on Metamath~\cite{10.48550/arxiv.2009.03393}, was ported to Lean by PACT~\cite{10.48550/arxiv.2102.06203}, and extended in the curriculum-learning system that followed~\cite{10.48550/arxiv.2202.01344}.
\texttt{LeanDojo}~\cite{10.48550/arxiv.2306.15626} is an open-source toolkit that extracts (state, tactic) training data from mathlib, bundled with a benchmark and \texttt{ReProver}, a retrieval-augmented tactic predictor. It is now the default environment for Lean ML work.
More recent infrastructure includes \texttt{Pantograph}~\cite{10.48550/arxiv.2410.16429}, which supports subgoal manipulation for tree-search methods, and \texttt{LeanInteract}~\cite{leaninteract}, a Python wrapper around the Lean 4 REPL. Editor-side tools such as \texttt{llmstep}~\cite{10.48550/arxiv.2310.18457} and \texttt{LeanCopilot}~\cite{10.48550/arxiv.2404.12534} surface LLM tactic suggestions to humans writing proofs interactively.

Reinforcement learning (RL) trains a model against a reward signal, updating its policy toward higher-reward outputs; in theorem proving, the reward is typically a binary success signal on completed proofs. RL-trained tactic policies are then paired with a tree-search procedure that expands candidate tactic steps and backtracks on failure to systematically explore the proof space. This combination was the first approach taken to automate Lean theorem proving. The line was opened by Polu et al.~\cite{10.48550/arxiv.2202.01344}, which applied expert-iteration-based RL on Lean through the \texttt{lean-gym} infrastructure.
Hypertree Proof Search~\cite{10.48550/arxiv.2205.11491} introduced a Monte Carlo Tree Search (MCTS)-inspired algorithm operating on proof hypergraphs. AlphaProof~\cite{Hubert_2025} extended this template with an AlphaZero-style value network; in combination with AlphaGeometry 2, it formed the system that reached silver-medal level at IMO 2024.
A subsequent open-source wave including Kimina-Prover~\cite{10.48550/arxiv.2504.11354}, DeepSeek-Prover-V2~\cite{10.48550/arxiv.2504.21801}, Goedel-Prover-V2~\cite{10.48550/arxiv.2508.03613}, and BFS-Prover~\cite{10.48550/arxiv.2502.03438,10.48550/arxiv.2509.06493} has continued the line of RL-trained-provers with stronger base models and richer training pipelines, sometimes replacing explicit search with model-internal reasoning.

A more recent contrasting approach is to take a general-purpose LLM and give it access to Lean tooling, without fine-tuning the base model.
The earliest published example is COPRA~\cite{10.48550/arxiv.2310.04353}, which used GPT-4 in-context with proof-environment feedback and lemma retrieval; most subsequent activity uses Anthropic's Model Context Protocol and the \texttt{Lean-LSP-MCP} bridge~\cite{lean_lsp_mcp}.
Recent systems include Ax-Prover, an MCP-native prover harness~\cite{10.48550/arxiv.2510.12787}; Numina-Lean-Agent, Claude Code with an extended Lean MCP~\cite{10.48550/arxiv.2601.14027}; and Delta Prover, an iterative-repair and decomposition agent~\cite{10.48550/arxiv.2507.15225}. These systems are competitive with specialized provers on standard benchmarks without additional training.

Subgoal decomposition splits a goal into separately-proved subproblems using Lean \texttt{sorry}s as placeholders, adopted by Hilbert~\cite{10.48550/arxiv.2509.22819}, Mechanic~\cite{10.48550/arxiv.2603.24465}, and Aristotle~\cite{10.48550/arxiv.2510.01346}.
APOLLO~\cite{10.48550/arxiv.2505.05758} uses Lean's structured error signals to localize failing subblocks in a generated proof, followed by a targeted repair pipeline.
Adaptive retrieval, with roots in \texttt{ReProver}~\cite{10.48550/arxiv.2306.15626}, surfaces relevant lemmas during proof search, used in REAL-Prover (over mathlib)~\cite{10.48550/arxiv.2505.20613} and LemmaHead (over olympiad textbooks)~\cite{10.48550/arxiv.2501.15797}.
Test-time reinforcement learning adapts the model to a specific problem at inference, applied by AlphaProof~\cite{Hubert_2025} and the Kimina-Prover TTRL extension~\cite{kimina_prover_ttrl_2025}.

\section{Benchmarks for LLM-Driven Formal Verification}

Most existing benchmarks for LLM-driven formal theorem proving are based on mathematics.
The de facto standard is miniF2F~\cite{10.48550/arxiv.2109.00110}, a set of olympiad-level math statements formalized in multiple proof assistants. Recent systems report increasingly high pass rates, with leading provers exceeding 90\%.
PutnamBench~\cite{10.48550/arxiv.2407.11214}, drawn from the undergraduate-level Putnam Competition, remains more challenging for current systems, with varying pass rates.
Leading systems are also evaluated on the latest IMO and Putnam contests to test performance on problems guaranteed to be outside training data.

Compared to math, code verification has far fewer benchmarks, many small or verifier-specific. DafnyBench~\cite{10.48550/arxiv.2406.08467} is the established Dafny benchmark, comprising 782 programs pairing specifications with code. The release baseline was a 68\% pass rate for Claude 3 Opus on filling in proof annotations; later evaluations report substantially higher pass rates.
For Lean and Verus, several benchmarks have been proposed since 2024 but none have been reused widely enough to dominate, leaving the field without a single standard reference.
\texttt{vericoding-benchmark}~\cite{10.48550/arxiv.2509.22908} is the largest code benchmark to date, evaluating end-to-end verified-code generation from formal specifications across more than 12,000 problems in Dafny, Verus, and Lean. At release, Lean verification was the hardest of the three with a reported pass rate of only 27\%, leaving substantial headroom for improvement via stronger base models, frameworks, or agentic scaffolds.

\chapter{Updated Baselines for Lean Verification}
\label{ch:baselines}

\section{Vericoding Benchmark Targeted Reproduction}
\label{sec:vericoder}

\subsection{Benchmark Subset Selection}

We begin by obtaining updated baselines for Lean verification on a targeted subset of high-quality verification tasks from \texttt{vericoding-benchmark}~\cite{10.48550/arxiv.2509.22908}, to track improvements in current models since the original release.

We evaluate on three Lean subsets: \texttt{bignum}, \texttt{verified\_cogen}, and \texttt{verina}, totaling 423 problems. The \texttt{bignum} subset comprises 62 tasks derived from arithmetic algorithms on big numbers commonly used in cryptography, written from scratch by the Vericoding authors in Dafny and translated into Lean using an LLM-based approach.
The \texttt{verina} subset is the original Lean release of the Verina benchmark from UC Berkeley's sunblaze group~\cite{10.48550/arxiv.2505.23135}, a curated collection of 189 modular tasks covering data structures, algorithms, and mathematical properties.
The \texttt{verified\_cogen} subset contains 172 tasks drawn from JetBrains Research's VerifiedCogen benchmark~\cite{jetbrains_verified_cogen_2025} of Verus programs focusing on memory safety and functional correctness, also translated via LLM.

We exclude the remaining Lean subsets in the benchmark (\texttt{NumpySimple}, \texttt{NumpyTriple}, \texttt{APPS-test}, \texttt{HumanEval/CLEVER}, \texttt{DafnyBench}, and \texttt{FVAPPS}) because their specifications were obtained either by autoformalization from natural language documentation, by LLM translation of informal Python sources, or in part from GitHub scrapes.
Among our three selected subsets, \texttt{bignum} was the most challenging in the original Vericoding evaluation with an overall pass rate of 12.9\% (model union over the nine evaluated LLMs), \texttt{verina} sat at an intermediate difficulty of 25.4\%, and \texttt{verified\_cogen} was the most solved at 44.2\%.

\subsection{Harness Setup}

We use the iterative self-correction harness of the original paper, referred to as the \emph{Vericoder}. At each iteration, the model is shown the original Lean spec, with \texttt{sorry} placeholders for missing implementations and \texttt{<vc-helpers>} tags for optional auxiliary definitions, and is asked to return a JSON array of textual replacements (one element per placeholder, in source order). These replacements are substituted into the spec and the resulting file is verified using the Lean compiler. If verification fails, the next iteration is given the failing code together with the Lean error log. Each new set of replacements is always inserted into the unmodified original spec rather than into the previous iteration's output. No external tools, retrieval, or web access is provided.
We keep the same proof-bypass validation step, which rejects any candidate whose generated blocks contain known cheating patterns, such as \texttt{sorry} or \texttt{admit} (Lean's incomplete-proof markers), \texttt{axiom} declarations, \texttt{unsafe} and \texttt{Unchecked.cast} type-correctness escape hatches, or \texttt{@[extern]} attributes that defer the implementation to native code.

Each spec is allowed five iterations. The original evaluation allowed ten attempts on the \texttt{bignum} subset and five on the other two; we use five attempts uniformly across all three subsets to keep per-spec compute comparable.
The paper does not note a specific reasoning configuration or an output token limit. We initially followed the code defaults when testing models: no explicit \texttt{reasoning\_effort} or thinking budget is passed to the API, and an output cap of 16{,}384 tokens per call. We ended up adopting a non-reasoning baseline due to issues with output token caps and for consistency of cross-model comparisons.

\subsection{Model Selection}

We evaluate ten models across US labs and leading open-source contenders. The closed-source set consists of seven models: GPT-5.4 and GPT-5.4-mini from OpenAI, Claude Sonnet 4.6 and Claude Haiku 4.5 from Anthropic, Gemini 3 Flash (Preview) and Gemini 3.1 Flash-Lite (Preview) from Google, and Grok 4.20 from xAI. All seven emit no reasoning at default API settings.

We excluded the following models from the closed-source set:
\begin{itemize}
	\item GPT-5.5: defaults to medium reasoning (can be turned off, not default); often spends entire output token cap on reasoning without returning content
	\item Claude Opus 4.7: defaults to no reasoning, but spontaneously emits \texttt{<think>} followed by raw chain-of-thought into visible output
	\item Gemini 3 Pro (Preview): defaults to high reasoning (cannot be turned off); also hits token cap
	\item Grok 4.3: defaults to low reasoning (can be turned off); skipped for consistency
\end{itemize}

The behavior of Opus 4.7 is particularly notable. Because the harness only extracts content within a JSON code block, the model is allowed to emit other prose without breaking the parser. Opus 4.7 appears to take advantage of this on almost every call, to a much greater extent than other models. Further, the output text appears to be a genuine model chain-of-thought, as opposed to a summary, with first-person hedging, false starts, and self-correction between failed proof attempts (Figure~\ref{fig:opus-think}).

In addition to consuming output tokens, this emission represents a leak in chain-of-thought protections implemented by US frontier labs in the last year. OpenAI introduced this practice with o1, surfacing only CoT summaries to users~\cite{openai_o1_system_card_2024}, and Anthropic and Google have since followed with redacted or rewritten reasoning blocks. These spontaneous \texttt{<think>} blocks in plain output circumvent that block on raw reasoning, and may be a means by which continued reasoning trace collection and distillation of frontier models can occur.

\begin{figure}[h]
\caption[Claude Opus 4.7 think-block excerpt on a \texttt{bignum} task.]{Excerpt of a Claude Opus 4.7 response on the \texttt{bignum\_Add\_NormalizeBitString} task at iteration 3. The model emits a \texttt{<think>} block of roughly 960 lines, including submission candidates, before closing the tag and producing a final JSON array. Total response length: 35{,}601 characters.}
\label{fig:opus-think}
\begin{verbatim}
<think>
Let me analyze the errors and fix them. The main issues are:

1. The `natToBits` helper has omega issues
2. The proof structure has indentation issues (`·` appears unexpected)
3. Need to prove `ValidBitString`

Let me simplify. I'll make `Add_` just return a valid string and prove
both parts.
...
[~960 lines of chain-of-thought omitted, including intermediate code blocks]
...
  sorry
```

Still stuck. OK submitting.
</think>

Looking at the errors, I need to fix:
1. The `omega` issues in `str2int_natToBits` ...

```json
["-- LLM HELPER\ndef natToBits : Nat -> List Char\n  | 0 => []\n  ...",
 "natToBitString (Str2Int s1 + Str2Int s2)",
 "refine <...>"]
```
\end{verbatim}
\end{figure}

We also evaluate the latest versions of three leading open-source models: GLM 5.1, DeepSeek V4 Pro, and Qwen 3.6 27B. GLM and DeepSeek were both evaluated in the original Vericoding paper~\cite{10.48550/arxiv.2509.22908}; we additionally include Qwen, which has recently developed into one of the strongest open-weight models~\cite{10.48550/arxiv.2505.09388}. These models all ship reasoning on by default, but since all seven closed-source models in our set run with reasoning off, we explicitly disable reasoning for these models to align baselines.

We run all models through OpenRouter. For closed models, we prefer providers with favorable pricing characteristics.\footnote{When available, we request flex processing, which offers a 50\% discount while accepting higher latency and lower throughput. This option is available for GPT models through OpenAI and Gemini models through Google Vertex; we configure OpenRouter to use those providers. This option is not available for Claude (we allow any provider), and Grok models (only available through the xAI provider). Even when requested, discounted processing is not guaranteed; adherence rates appear to vary with global inference demand, with best results outside US business hours.} For open models, we pin to official or non-quantized providers when possible.\footnote{We use the official providers for GLM 5.1 (Z.ai) and Qwen 3.6 27B (Alibaba), which do not report any quantization; model weights are published online as BF16. The official DeepSeek provider is blocked due to data policy guardrails; we use third-party providers for DeepSeek V4 Pro which report a FP8 quantization.}

\subsection{Results}

Table~\ref{tab:lean-leaderboard} reports per-subset and aggregate solve rates, theoretical cost (at published provider rates), and total tokens (input + output). GPT-5.4 leads in non-reasoning Lean verification at 24.6\%. Claude Sonnet 4.6 and Gemini 3 Flash form a second tier above 10\%, followed by the three open-weight models. Claude Haiku 4.5 solves four specs, while GPT-5.4-mini, Gemini 3.1 Flash-Lite and Grok 4.20 fail to solve any of the 423 tasks.

To compare against the original Vericoding evaluation, we recompute the paper's per-model pass rates restricted to the same three subsets we ran (Table~\ref{tab:cross-generation}). Generational deltas are modest, with the closed-source US models gaining between +0.7\% and +4.0\% over previous generation siblings. At the budget tier the picture reverses: GPT-5.4-mini, Gemini 3.1 Flash-Lite, and Grok 4.20 all regressed to 0\% from nonzero paper-baseline numbers. This pattern points to very little capability improvement on mid- and budget-tier models when reasoning is disabled. The current generation's effort appears to have been concentrated on reasoning-mode behavior, which this non-reasoning baseline does not capture.

A change from the previous year is that the open-weight frontier now outperforms the smallest closed models without reasoning: GLM 5.1 (4.3\%), DeepSeek V4 Pro (2.1\%), and Qwen 3.6 27B (1.7\%) all sit above Claude Haiku 4.5 (0.9\%). Compared to previous generations, both GLM 5.1 (+3.3\% over GLM 4.5) and DeepSeek V4 Pro (+2.1\% over DeepSeek V3.1) show improvements, though the absolute gap to GPT-5.4 remains large.

\begin{table}
\caption[Cross-model Vericoder reproduction results on the Lean 423-spec set.]{Cross-model reproduction of the Vericoding Lean benchmark, non-reasoning baseline. Each model is given five iterations per spec, matching the paper's Lean protocol~\cite{10.48550/arxiv.2509.22908}, except on \texttt{bignum}. Closed-source rows run with reasoning off by default. Open-source rows (GLM, DeepSeek, Qwen) have reasoning explicitly disabled. The \emph{model union} counts a task as solved if any of the ten models solved it. \emph{Cost} is computed at provider published list rates; \emph{Tokens} is the sum of input and output.}
\label{tab:lean-leaderboard}
\renewcommand{\arraystretch}{1.15}
\centering
\scalebox{0.9}{
\begin{tabular}{l r r r >{\columncolor{gray!10}\bfseries}r r r}
\toprule
 & \textbf{BigNum} & \textbf{VerifCogen} & \textbf{Verina} & \textbf{Totals $\downarrow$} & \textbf{Cost} & \textbf{Tokens} \\
\midrule
\textbf{Lean}                 & 62 tasks & 172 tasks & 189 tasks & \textbf{423 tasks} & --     & --    \\
\midrule
\textbf{gpt-5.4}              &  9.7\% & 33.1\% & 21.7\% & 24.6\% & \$23.57 & 5.2M \\
\textbf{claude-sonnet-4.6}    &  9.7\% & 15.7\% & 15.3\% & 14.7\% & \$50.98 & 7.8M \\
\textbf{gemini-3-flash}       &  0.0\% &  8.7\% & 14.8\% & 10.2\% & \$5.71  & 6.4M \\
\textbf{glm-5.1}              &  0.0\% &  4.1\% &  5.8\% &  4.3\% & \$15.82 & 8.0M \\
\textbf{deepseek-v4-pro}      &  0.0\% &  2.9\% &  2.1\% &  2.1\% & \$17.80 & 6.9M \\
\textbf{qwen-3.6-27b}         &  0.0\% &  2.9\% &  1.1\% &  1.7\% & \$8.56  & 6.0M \\
\textbf{claude-haiku-4.5}     &  0.0\% &  1.7\% &  0.5\% &  0.9\% & \$12.36 & 8.3M \\
\textbf{gpt-5.4-mini}         &  0.0\% &  0.0\% &  0.0\% &  0.0\% & \$3.57  & 3.4M \\
\textbf{gemini-3.1-flash-lite}  &  0.0\% &  0.0\% &  0.0\% &  0.0\% & \$1.72  & 5.8M \\
\textbf{grok-4.20}            &  0.0\% &  0.0\% &  0.0\% &  0.0\% & \$5.66  & 4.4M \\
\midrule
\cellcolor{gray!10}\textbf{model union} &
\cellcolor{gray!10}\bfseries 12.9\% &
\cellcolor{gray!10}\bfseries 42.4\% &
\cellcolor{gray!10}\bfseries 29.1\% &
\cellcolor{gray!40}\bfseries 32.2\% &
\cellcolor{gray!10}-- &
\cellcolor{gray!10}-- \\
\bottomrule
\end{tabular}
}
\end{table}

\begin{table}
\caption[Reproduction vs.\ Vericoding paper baseline, comparing model generations.]{Comparison of our reproduction against the Vericoding paper baseline~\cite{10.48550/arxiv.2509.22908}, both restricted to the same three Lean subsets (\texttt{bignum}, \texttt{verified\_cogen}, \texttt{verina}; $n = 423$). Paper percentages recomputed on the three tested subsets. Reproduction deviates slightly as \texttt{bignum} also has five iterations in the reproduction, as opposed to ten in the original paper. \emph{Change} computed from unrounded pass rates.}
\label{tab:cross-generation}
\renewcommand{\arraystretch}{1.15}
\centering
\begin{tabular}{l r l r r}
\toprule
\textbf{Our model} & \textbf{Ours} & \textbf{Paper sibling} & \textbf{Paper} & \textbf{Change} \\
\midrule
\textbf{gpt-5.4}              & 24.6\% & gpt-5              & 22.5\% & +2.1\% \\
\textbf{claude-sonnet-4.6}    & 14.7\% & claude-sonnet-4    & 10.6\% & +4.0\% \\
\textbf{gemini-3-flash}       & 10.2\% & gemini-2.5-pro     &  9.5\% & +0.7\% \\
\textbf{glm-5.1}              &  4.3\% & glm-4.5            &  0.9\% & +3.3\% \\
\textbf{deepseek-v4-pro}      &  2.1\% & deepseek-chat-v3.1 &  0.0\% & +2.1\% \\
\textbf{qwen-3.6-27b}         &  1.7\% & \textit{no sibling} & ---  & --- \\
\textbf{claude-haiku-4.5}     &  0.9\% & \textit{no sibling} & ---  & --- \\
\textbf{gpt-5.4-mini}         &  0.0\% & gpt-5-mini         &  2.6\% & $-2.6$\% \\
\textbf{gemini-3.1-flash-lite}  &  0.0\% & gemini-2.5-flash   &  0.2\% & $-0.2$\% \\
\textbf{grok-4.20}            &  0.0\% & grok-code          &  2.4\% & $-2.4$\% \\
\bottomrule
\end{tabular}
\end{table}

\subsection{Chain-of-Thought Emission}

Looking more closely at model responses, several models also emit visible chain-of-thought into the output channel. The Vericoder prompt instructs the model to output JSON, but does not specifically forbid prose: it instructs the model to ``Return a JSON array with EXACTLY $N$ replacements\dots in order from top to bottom''.

We analyze the distribution and proportion of visible thinking across models, prorating each spec's response characters against its API-reported \texttt{output\_tokens}. Results are reported in Table~\ref{tab:cot-prose-audit}.
Anthropic models spend a large share of every output budget on visible deliberation before their final JSON: Claude Sonnet 4.6 averages $56.6\%$, and Claude Haiku 4.5 averages $26.9\%$. The open-source models, asked to disable reasoning, emit qualitatively similar but milder deliberation: Qwen 3.6 27B at $23.7\%$ of budget, DeepSeek V4 Pro at $13.6\%$, and GLM 5.1 at $10.4\%$.
Grok, Gemini, and GPT models emit little to no visible thinking.

Further, many responses contain multiple JSON code blocks in succession (a draft, then a self-corrected refinement, then another); however, the harness regex extracts only the first. We measure and report rates of multiple code blocks in Table~\ref{tab:multi-block}. This behavior concentrates in the same models that emit visible deliberation: Claude Sonnet 4.6 emits multiple JSON blocks on $53.9\%$ of its $1{,}817$ responses (worst-case 13 blocks spanning $41{,}082$ characters), Claude Haiku 4.5 on $16.0\%$, and the open-weight models at lower rates (Qwen $8.3\%$, DeepSeek $1.1\%$, GLM $0.7\%$). GPT-5.4, GPT-5.4-mini, and Gemini 3.1 Flash-Lite always emit a single code block.

The code-block-capturing logic of the Vericoding harness appears designed for small amounts of incidental text (such as ``Here's the answer'' or similar), not extended within-response reasoning with successive candidate blocks.
The cross-provider comparison is therefore unfair in two opposing directions for the Anthropic and open-source models. They may gain an advantage from extended within-response reasoning that refines their answers, but are simultaneously penalized when that reasoning emits intermediate JSON code blocks. The harness first-match extractor consumes the first generated code block as final, discarding any subsequent refinements. As a result, cross-model comparisons are not strictly comparable due to differing model behavior and handling.
Future reproductions of the Vericoding harness could either adjust the original extraction logic to handle multiple code blocks, or explicitly instruct models to emit only a single code block with no visible deliberation.

\begin{table}
\caption[Proportion of output emitted as thinking, for each model.]{Proportion of output emitted as thinking, for each model. Token counts prorate output characters against API-reported \texttt{output\_tokens} for each model. The model output cap is set to $16{,}384$ tokens.}
\label{tab:cot-prose-audit}
\renewcommand{\arraystretch}{1.15}
\centering
\scalebox{0.9}{
\begin{tabular}{l r r r r r}
\toprule
\textbf{Model} & \textbf{N} & \textbf{prose / total} & \textbf{mean tokens} & \textbf{p95 tokens} & \textbf{max tokens} \\
\midrule
\textbf{claude-sonnet-4.6}    & 1{,}817 & 56.6\% & 717 & 1{,}439 & 15{,}595 \\
\textbf{claude-haiku-4.5}     & 2{,}010 & 26.9\% & 180 & 889     & 2{,}222  \\
\textbf{qwen-3.6-27b}         & 1{,}633 & 23.7\% & 237 & 1{,}646 & 13{,}538 \\
\textbf{deepseek-v4-pro}      & 1{,}774 & 13.6\% & 111 & 239     & 15{,}620 \\
\textbf{glm-5.1}              & 2{,}028 & 10.4\% & 97  & 253     & 13{,}245 \\
\textbf{grok-4.20}            & 1{,}597 & 2.3\%  & 8   & 0       & 9{,}724  \\
\textbf{gemini-3-flash}       & 1{,}927 & 0.1\%  & 0   & 0       & 411      \\
\textbf{gemini-3.1-flash-lite}  & 2{,}032 & 0.0\%  & 0   & 0       & 160      \\
\textbf{gpt-5.4}              & 1{,}768 & 0.0\%  & 0   & 0       & 0        \\
\textbf{gpt-5.4-mini}         & 1{,}584 & 0.0\%  & 0   & 0       & 0        \\
\bottomrule
\end{tabular}
}
\end{table}

\begin{table}
\caption[Multi-block emission rates, by model.]{Multi-block emission: fraction of responses that contain more than one code block. \emph{multi\%} is the percentage of responses with more than one block; \emph{med.\ blocks} is the median block count among multi-block responses.}
\label{tab:multi-block}
\renewcommand{\arraystretch}{1.15}
\centering
\scalebox{0.9}{
\begin{tabular}{l r r r r}
\toprule
\textbf{Model} & \textbf{N} & \textbf{multi\%} & \textbf{med. blocks} & \textbf{max blocks} \\
\midrule
\textbf{claude-sonnet-4.6}    & 1{,}817 & 53.9\% & 2 & 13 \\
\textbf{claude-haiku-4.5}     & 2{,}010 & 16.0\% & 2 & 5  \\
\textbf{qwen-3.6-27b}         & 1{,}633 & 8.3\%  & 3 & 13 \\
\textbf{deepseek-v4-pro}      & 1{,}774 & 1.1\%  & 3 & 15 \\
\textbf{glm-5.1}              & 2{,}028 & 0.7\%  & 2 & 10 \\
\textbf{grok-4.20}            & 1{,}597 & 0.1\%  & 2 & 2  \\
\textbf{gemini-3-flash}       & 1{,}927 & 0.1\%  & 2 & 2  \\
\textbf{gpt-5.4}              & 1{,}768 & 0.0\%  & — & —  \\
\textbf{gemini-3.1-flash-lite}  & 2{,}032 & 0.0\%  & — & —  \\
\textbf{gpt-5.4-mini}         & 1{,}584 & 0.0\%  & — & —  \\
\bottomrule
\end{tabular}
}
\end{table}

\section{Agentic Loop Baseline}
\label{sec:agent-baseline}

\subsection{Motivation}

In the previous section, we reran the original Vericoding harness with updated models. However, this does not represent an honest baseline in the context of current model capabilities, as the configuration does not make use of three methodologies that are commonly used in model evaluations, and known to improve performance.

The first is reasoning. Without it, the model must begin decoding its answer immediately upon receiving the prompt, without any room for planning or outlining. Modern frontier models are post-trained for this mode~\cite{10.48550/arxiv.2501.12948,openai_o1_system_card_2024} and tend to use it naturally when permitted.

The second is the agentic loop. Instead of starting every round with a new context, each round's submission and result are appended as new turns to the end of the conversation.
The model can refine its approach across rounds by building on prior attempts, as opposed to being forced to rederive understanding each turn.
Model providers optimize for this mode through server-side KV caching, eliminating the forward pass over the prior conversation on continuing turns.

The third is access to external knowledge via tools. In the context of Lean, this takes the form of a Lean MCP server, which provides access to a search tool. The tool provides access to multiple Lean search providers which handle natural language and type-pattern queries, allowing the model to broadly access lemma names and confirm type signatures when constructing proofs.

Almost all current LLM-driven Lean theorem proving systems use one or more of these methods, and recent agent-loop harnesses like Numina-Lean-Agent~\cite{10.48550/arxiv.2601.14027} combine all three, layered with additional scaffolding.
For a modern, capability-honest baseline, we build a harness combining an agent loop, reasoning, and Lean search, against which tree-search algorithms in later sections can be evaluated.

\subsection{Experimental Setup}

We maintain the same input and output format as the original Vericoding setup: each spec is presented with one or more \texttt{sorry} holes, and submissions are expressed as a JSON array of replacement strings. We change the submission method from text output to a \texttt{submit\_code} tool, which enforces the replacements JSON formatting via schema validation rather than relying on the model to produce a valid code block. We also filter the compiler output to remove build noise before returning feedback as tool output. We expose the \texttt{search\_mathlib} tool from the \texttt{Lean-LSP-MCP} server~\cite{lean_lsp_mcp}, which dispatches to LeanFinder and Loogle; the model can choose which provider to use per call.

As the model now has a search tool in addition to submission, we initially set the budget for each spec to ten LLM calls. The loop terminates early on a \texttt{submit\_code} call that verifies, otherwise running until the budget is exhausted. If the model outputs text instead of a tool call, it is nudged to continue calling tools.
We evaluate across the six models from OpenAI, Anthropic, and Google tested in the previous section. Reasoning effort is explicitly set to medium across all models, and we use the same 423-spec subset.

\subsection{Initial Results}

With ten LLM calls in the agent harness and medium reasoning, the union over all six models solves 263 / 423 specs (62.2\%). Table~\ref{tab:agent-k10-headline} reports per-subset and aggregate solve rates.

The six models all improve substantially over the Vericoder baseline of Section~\ref{sec:vericoder} on the same 423 specs: GPT-5.4 lifts from 24.6\% to 60.3\% (a 2.45$\times$ gain), while GPT-5.4-mini, Claude Haiku 4.5, and Gemini 3.1 Flash-Lite score significantly more than the previous 0-1\%. A small number of runs (1.8\% of spec and model combinations) crash due to output-truncation errors when reasoning tokens consume the entire per-call output budget, with 74\% of crashes occurring on Claude Sonnet 4.6 runs.

For a fair comparison between agent and Vericoder pipelines, we compare token spend and solve rate by round for the agent harness. As the agentic loop relies on the reuse of a cached input prefix across LLM calls, we count \emph{unique tokens}: the sum of input and output tokens across all unique blocks, not counting duplicate cache reads. We compute cost by multiplying unique input and output tokens by respective billing rates, which corresponds to the assumption of optimal caching for all input blocks and free cache reads.\footnote{In practice, actual cache hit rates vary depending on timing of requests, provider routing, and other uncontrollable factors; however, they are generally high for a simple agentic loop as described. Cached tokens are also not entirely free; they are typically discounted around 90\% for recent models. For a 10-turn agentic loop, 10 cache reads averaging half the input adds approximately 50\% to the input price. Some providers, such as Anthropic, additionally charge extra for cache writes over standard input prices.} The original Vericoder constructs a new prompt each turn, resulting in minimal caching across rounds.\footnote{The Vericoder prompt is not optimized for prefix caching, as it increments the turn count early in the prompt. We still observe a small amount of prompt caching for providers that automatically cache within blocks. The prompt consists of instructions + error + original file + last submitted file; with a bit of reordering, the instructions + original file could be cached, reducing unique input by around 50\%. However, even a 50\% decrease in Vericoder unique tokens would not significantly change the results at matched tokens.}

\begin{table}
\caption[Per-subset solve rates of the agent baseline at $K=10$.]{Per-subset solve rates at $K=10$ across the six LLMs evaluated, alongside the six-model union and the Vericoder union for the same six models. \emph{Cost} is total spend at provider list rates assuming perfect prompt caching and free cache reads; \emph{Tokens} is the corresponding unique token count (output includes reasoning).}
\label{tab:agent-k10-headline}
\renewcommand{\arraystretch}{1.15}
\centering
\scalebox{0.9}{
\begin{tabular}{l r r r >{\columncolor{gray!10}\bfseries}r r r}
\toprule
 & \textbf{BigNum} & \textbf{VerifCogen} & \textbf{Verina} & \textbf{Totals $\downarrow$} & \textbf{Cost} & \textbf{Tokens} \\
\midrule
\textbf{Lean}                 & 62 tasks & 172 tasks & 189 tasks & \textbf{423 tasks} & --       & --    \\
\midrule
\textbf{gpt-5.4}              & 19.4\% & 75.0\% & 60.3\% & 60.3\% &  \$65.06 & 10.9M \\
\textbf{gpt-5.4-mini}         & 12.9\% & 43.6\% & 38.1\% & 36.6\% &  \$25.88 & 11.4M \\
\textbf{claude-sonnet-4.6}    & 11.3\% & 34.9\% & 30.7\% & 29.6\% &  \$41.10 &  5.9M \\
\textbf{gemini-3-flash}       &  6.5\% & 25.6\% & 24.9\% & 22.5\% &  \$16.42 &  6.3M \\
\textbf{claude-haiku-4.5}     & 11.3\% & 24.4\% & 17.5\% & 19.4\% &   \$9.54 &  4.4M \\
\textbf{gemini-3.1-flash-lite}  &  3.2\% & 21.5\% & 11.1\% & 14.2\% &   \$2.23 &  2.7M \\
\midrule
\cellcolor{gray!10}\textbf{model union} &
\cellcolor{gray!10}\bfseries 19.4\% &
\cellcolor{gray!10}\bfseries 78.5\% &
\cellcolor{gray!10}\bfseries 61.4\% &
\cellcolor{gray!40}\bfseries 62.2\% &
\cellcolor{gray!10}-- &
\cellcolor{gray!10}-- \\
\textbf{Vericoder (6 models)} &
12.9\% &
41.9\% &
28.6\% &
\cellcolor{gray!10}31.7\% &
-- &
-- \\
\bottomrule
\end{tabular}
}
\end{table}

\begin{figure}
\centering
\includegraphics[width=0.95\textwidth]{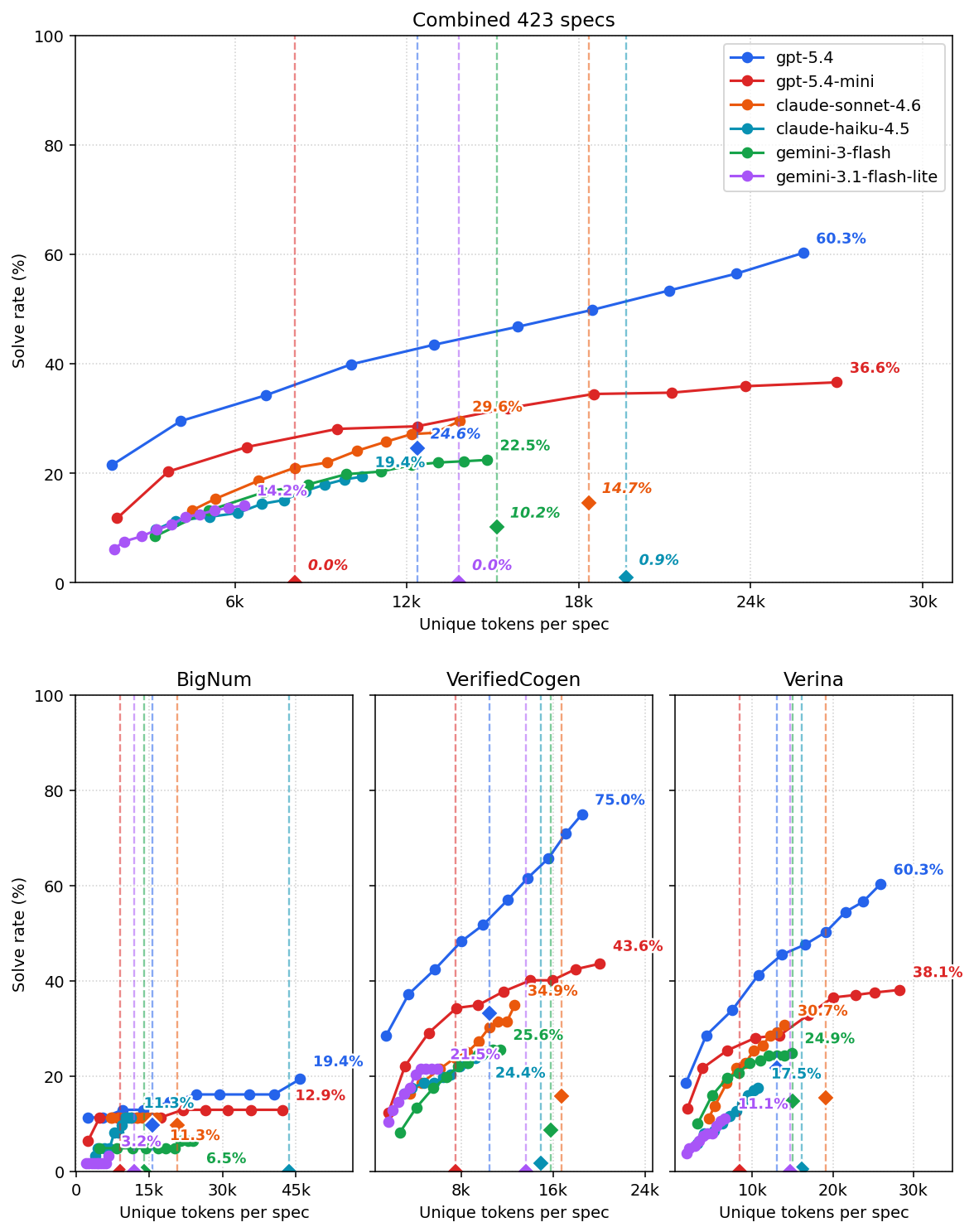}
\caption[$K=10$ agent solve rate vs.\ unique tokens, full set and per subset.]{Solve rate vs.\ average unique tokens per spec, iterated over $K = 1, 2, \dots, 10$ model calls, for each of the six LLMs. \emph{Top}: full 423-spec set. \emph{Bottom row, left to right}: \texttt{bignum}, \texttt{verified\_cogen}, and \texttt{verina} subsets. Dashed vertical lines mark each LLM's Vericoder tokens per spec from Section~\ref{sec:vericoder}, with diamond markers at the corresponding Vericoder solve rate.}
\label{fig:pareto-compare}
\end{figure}

Figure~\ref{fig:pareto-compare} plots solve rate against unique tokens for the six models, iterated over the number of model calls. Original token count and solve rate from Vericoder are also indicated.
At matched unique-token budgets, the OpenAI, Anthropic, and Gemini families all lie above their Vericoder counterparts from Section~\ref{sec:vericoder}.
Gemini 3.1 Flash-Lite emits the fewest tokens of any model evaluated, and at matched unique-token budgets it outperforms Gemini 3 Flash.
GPT-5.4 emits fewer cumulative tokens than GPT-5.4-mini despite solving more specs. Both OpenAI models use adaptive reasoning, which allocates reasoning tokens based on problem difficulty. GPT-5.4 can resolve specs in fewer tokens because the problem is easier relative to its capabilities, whereas GPT-5.4-mini thinks longer on the same problem without always succeeding.
Claude Sonnet 4.6 uses the fewest tokens per spec of the mid-tier models, but performs worse than the budget-tier GPT-5.4-mini at a higher cost.

The bottom row of Figure~\ref{fig:pareto-compare} shows per-subset solve trajectories. \texttt{bignum} remains challenging across all six models, with the six-model union covering only 12 / 62 specs (19.4\%) versus 62.2\% on the full set. This union is exactly the set of specs solved by GPT-5.4, with every other model's solves being a strict subset.
The two Claude models tie at 11.3\% (7 / 62 each), while both Gemini models perform particularly poorly on this subset, with Gemini 3 Flash leveling off at 6.5\% of specs solved and Gemini 3.1 Flash-Lite only reaching 3.2\%.

\subsection{Agent Tool Call Behavior}
\label{subsec:agent-tools}

\begin{figure}
\centering
\includegraphics[width=\textwidth]{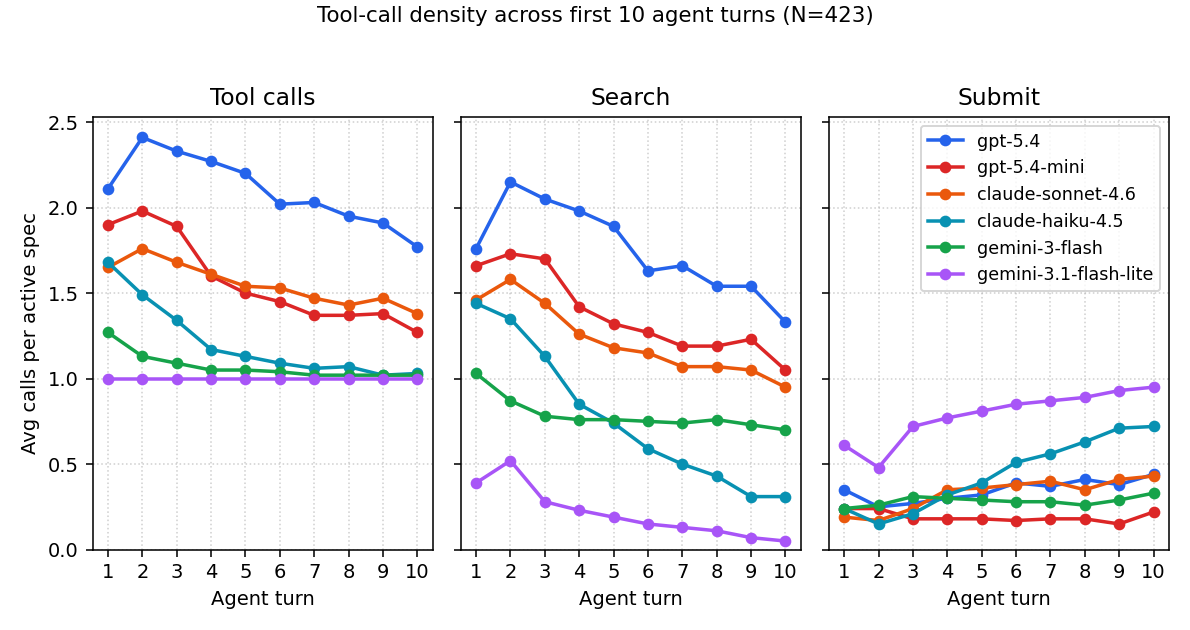}
\caption[Per-turn tool-call rates across the first ten agent turns.]{Average total tool calls per turn (left), \texttt{search\_mathlib} calls per turn (middle), and \texttt{submit\_code} calls per turn (right), across the first ten turns of the agent loop on the full 423-spec set. Per-turn averages are conditional on a spec still being active at that round.}
\label{fig:tool-choice-per-turn}
\end{figure}

For a comparison of submission counts, we look into tool-calling behavior of models in the agent harness. Figure~\ref{fig:tool-choice-per-turn} plots per-turn averages of total tool calls, \texttt{search\_mathlib} calls, and \texttt{submit\_code} calls across the first ten turns, normalized by the number of specs still active at that round.

Most models batch multiple tool calls, more in the initial rounds than later ones. OpenAI models use parallel tool calling most aggressively, with GPT-5.4 issuing three \texttt{search\_mathlib} calls in 77\% of search turns, with a maximum of four.
Claude Sonnet 4.6 calls two parallel searches on 82\% of its search turns and never more than two.
Gemini 3 Flash only requests multiple searches on 8\% of search turns, and Gemini 3.1 Flash-Lite never batches, always making one tool call per turn.
No model ever issues more than one \texttt{submit\_code} call in a single turn, though parallel submissions are not explicitly forbidden. Mixing a search and a submission inside one turn is also rare: only Claude Sonnet 4.6 (1.4\% of turns) and Gemini 3 Flash (0.14\%) ever do so, while the other four models always keep the two tools separate.

Table~\ref{tab:tool-choice-summary} aggregates per-turn averages over the ten-turn window and the search-to-submit ratio for each model.
The OpenAI models utilize retrieval most heavily: GPT-5.4-mini issues 7.2 searches for every submit and GPT-5.4 issues 5.3. GPT-5.4-mini's higher ratio likely reflects the problems being harder relative to its capabilities, requiring more lemma lookups before each submission attempt. Conversely, weaker-performing models have a lower search-to-submit ratio, such as Claude Haiku 4.5 with a ratio of 1.8. Gemini 3.1 Flash-Lite inverts the ratio entirely at 0.28, submitting in most turns as opposed to searching. This suggests the weaker models are less tuned to the task's combined demands of logical reasoning and lemma lookup.

For every model except Gemini 3.1 Flash-Lite, the sum of average submit calls per turn is less than five, indicating that models on average invoke the Lean compiler fewer times than in the Vericoder baseline.

\begin{table}
\caption[Per-model tool-call behavior across ten agent turns.]{Per-model summary of tool-call behavior across the first ten agent turns on the full 423-spec set. Totals sum per-turn averages over rounds 1--10; \emph{Ratio} is total \texttt{search\_mathlib} divided by total \texttt{submit\_code}.}
\label{tab:tool-choice-summary}
\renewcommand{\arraystretch}{1.15}
\centering
\begin{tabular}{l r r r r}
\toprule
 & \textbf{Total search} & \textbf{Total submit} & \textbf{Total calls} & \textbf{Ratio} \\
\midrule
\textbf{gpt-5.4}              & 17.53 & 3.48 & 21.00 & 5.29 \\
\textbf{gpt-5.4-mini}         & 13.77 & 1.93 & 15.70 & 7.17 \\
\textbf{claude-sonnet-4.6}    & 12.24 & 3.28 & 15.52 & 3.91 \\
\textbf{claude-haiku-4.5}     &  7.63 & 4.46 & 12.09 & 1.80 \\
\textbf{gemini-3-flash}       &  7.86 & 2.83 & 10.70 & 2.81 \\
\textbf{gemini-3.1-flash-lite}  &  2.11 & 7.89 & 10.00 & 0.28 \\
\bottomrule
\end{tabular}
\end{table}

\subsection{Longer-Context Evaluations}
\label{subsec:longer-context}

The Figure~\ref{fig:pareto-compare} curves climb roughly linearly through $K = 10$ for most models. GPT-5.4-mini solves a greater proportion of specs in the first few turns, then marginal solve rate starts decreasing. GPT-5.4 shows no sign of slowing, with a similar number of additional solves each turn.

Test-time scaling theory predicts that solve rate eventually flattens: as a model approaches its capability ceiling for the task, additional tokens provide less additional benefit, and the curve plateaus. The $K = 10$ trajectories give little to no evidence of this transition, indicating that more turns may continue to provide increases in overall benchmark performance.

To test further scaling behavior, we extend the agent loop on the same 423-spec set past $K = 10$. We narrow our tests to OpenAI and Google models, as the OpenAI models produced the highest absolute solve rates at $K = 10$, the Google models offer a low-cost comparison, and both are offered by providers with favorable pricing. We resume every spec and model in increments of ten LLM calls, ending at a final budget of $K = 50$ per spec.

\begin{figure}
\centering
\includegraphics[width=0.95\textwidth]{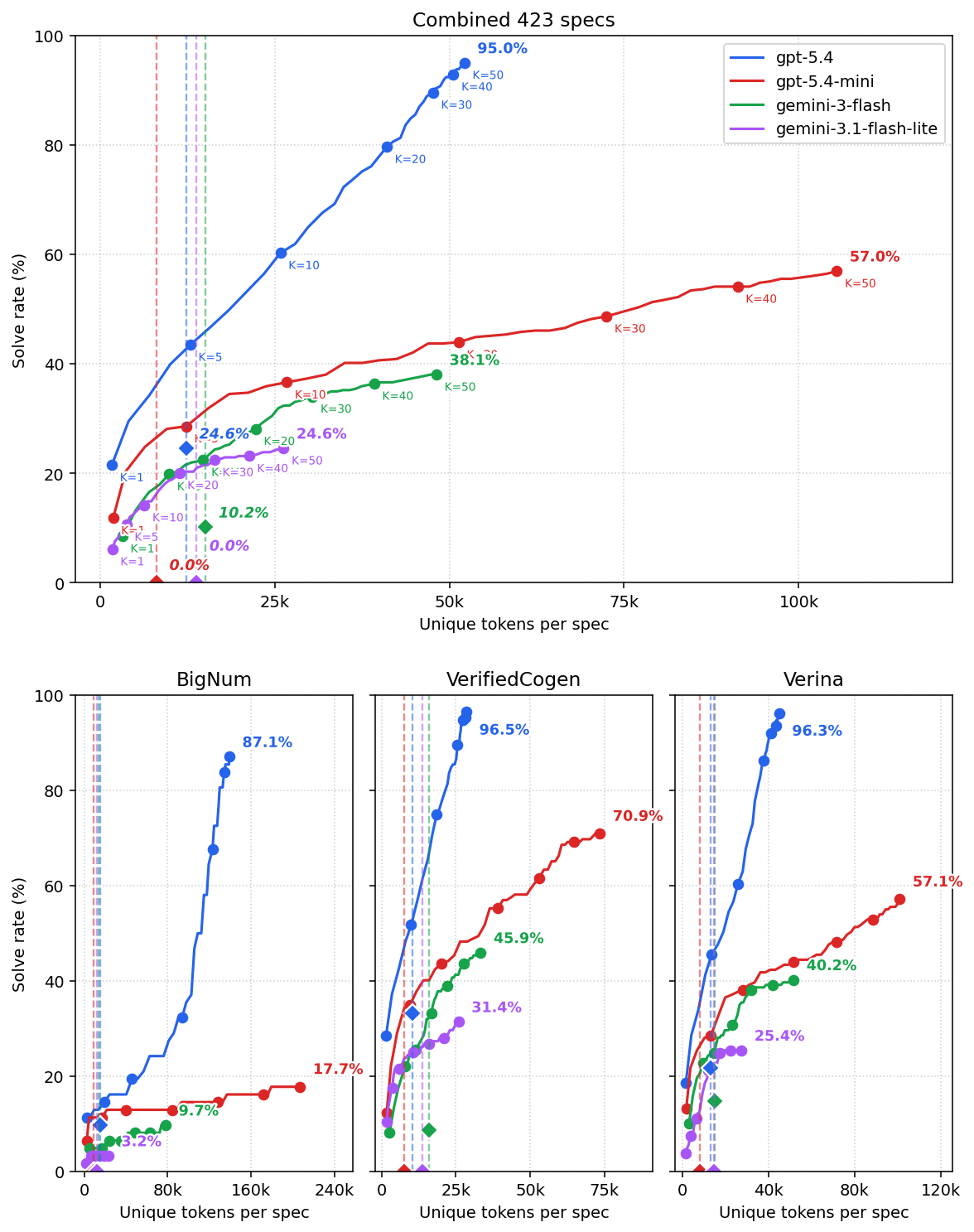}
\caption[$K=50$ agent solve rate vs.\ unique tokens, full set and per subset.]{$K=50$ continuation of four-LLM agent comparison, beyond the $K=10$ cap of Figure~\ref{fig:pareto-compare}. \emph{Top}: full 423-spec set. \emph{Bottom row, left to right}: \texttt{bignum}, \texttt{verified\_cogen}, and \texttt{verina} subsets. Dashed vertical lines mark each LLM's Vericoder tokens per spec from Section~\ref{sec:vericoder}, with diamond markers at the corresponding Vericoder solve rate.}
\label{fig:pareto-compare-extended}
\end{figure}

Figure~\ref{fig:pareto-compare-extended} shows pass rate curves through $K = 50$ for all four models. GPT-5.4 climbs from 60.3\% at $K = 10$ to 95.0\% at $K = 50$, while the other three models each improve by roughly 10 to 20 percentage points over the same range.

Table~\ref{tab:agent-extended-kgrid} reports per-subset pass rates and per-spec tokens up to $K=50$. GPT-5.4 reaches 87.1\% on \texttt{bignum}, 96.5\% on \texttt{verified\_cogen}, and 96.3\% on \texttt{verina}; the three weaker models pick up most of their additional solves on \texttt{verified\_cogen} and \texttt{verina} while making little to no additional progress on \texttt{bignum}.

All models except Gemini 3.1 Flash-Lite spend the fewest tokens on \texttt{verified\_cogen} and substantially more tokens on \texttt{bignum}, reflecting relative subset difficulties and adaptive reasoning effort; Gemini 3.1 Flash-Lite does not show appreciable output length differences across subsets. For GPT-5.4 and to a lesser extent GPT-5.4-mini, marginal tokens per round decreases as $K$ grows, since specs solved at earlier rounds no longer contribute additional tokens to the running average.

Particularly notable is the solve rate of GPT-5.4 on \texttt{bignum} across rounds, shown in Table~\ref{tab:gpt-bignum-kcurve}. After an initial nine solves at $K = 5$, marginal progress slows to three to five additional specs per five-round increment, reaching 20 / 62 (32.3\%) at $K = 20$. The curve then accelerates sharply: 22 specs are solved between $K = 20$ to $30$, more than doubling the cumulative solve count. The model solves another eight specs to reach 50 / 62 (80.6\%) at $K = 35$ before the rate decays, with only four more solves by $K = 50$. Per-task solve trajectories of this shape have been noted in the test-time scaling literature~\cite{10.48550/arxiv.2407.21787}.

The 95.0\% pass rate at $K = 50$ for GPT-5.4 indicates that \texttt{vericoding-benchmark} is effectively saturated by recent model capabilities. GPT-5.4 is currently considered the mid-tier model in OpenAI's lineup, and saturation is achieved at medium reasoning; it is possible that frontier models such as GPT-5.5, with higher reasoning effort, can reach saturation in significantly fewer turns.

Taken together, the reasoning, agentic loop, and MCP tool changes produce substantial improvements over the Vericoder baseline, across subsets and models. The agent harness establishes a strong baseline reflecting modern model capabilities, and the Pareto curves provide benchmarks against which tree-search methods can be evaluated at matched token usage.

\begin{table}
\caption[Agent pass rate at varying $K$, by subset.]{Agent benchmark pass rate at varying LLM-call budgets $K$, across the four LLMs evaluated. Each cell shows pass rate~/ average unique tokens per spec at that $K$.}
\label{tab:agent-extended-kgrid}
\renewcommand{\arraystretch}{1.15}
\centering
\scalebox{0.85}{
\begin{tabular}{l r r r r r >{\columncolor{gray!10}\bfseries}r}
\toprule
\textbf{Subset} & \textbf{n} & \textbf{$K{=}10$} & \textbf{$K{=}20$} & \textbf{$K{=}30$} & \textbf{$K{=}40$} & \textbf{$K{=}50$} \\
\midrule
\multicolumn{7}{l}{\textbf{gpt-5.4}} \\
\midrule
BigNum     &  62 & 19.4\% / 46k & 32.3\% / 94k & 67.7\% / 123k & 83.9\% / 135k & 87.1\% / 139k \\
VerifCogen & 172 & 75.0\% / 19k & 89.5\% / 26k & 94.8\% / 27k  & 95.3\% / 28k  & 96.5\% / 29k \\
Verina     & 189 & 60.3\% / 26k & 86.2\% / 38k & 92.1\% / 41k  & 93.7\% / 44k  & 96.3\% / 45k \\
\cellcolor{gray!10}\textbf{Combined} &
\cellcolor{gray!10}\bfseries 423 &
\cellcolor{gray!10}\bfseries 60.3\% / 26k &
\cellcolor{gray!10}\bfseries 79.7\% / 41k &
\cellcolor{gray!10}\bfseries 89.6\% / 48k &
\cellcolor{gray!10}\bfseries 92.9\% / 51k &
\cellcolor{gray!40}\bfseries 95.0\% / 52k \\
\midrule
\multicolumn{7}{l}{\textbf{gpt-5.4-mini}} \\
\midrule
BigNum     &  62 & 12.9\% / 40k & 12.9\% / 84k & 14.5\% / 129k & 16.1\% / 172k & 17.7\% / 207k \\
VerifCogen & 172 & 43.6\% / 20k & 55.2\% / 39k & 61.6\% / 53k  & 69.2\% / 65k  & 70.9\% / 74k \\
Verina     & 189 & 38.1\% / 28k & 43.9\% / 52k & 48.1\% / 72k  & 52.9\% / 89k  & 57.1\% / 101k \\
\cellcolor{gray!10}\textbf{Combined} &
\cellcolor{gray!10}\bfseries 423 &
\cellcolor{gray!10}\bfseries 36.6\% / 27k &
\cellcolor{gray!10}\bfseries 44.0\% / 51k &
\cellcolor{gray!10}\bfseries 48.7\% / 72k &
\cellcolor{gray!10}\bfseries 54.1\% / 91k &
\cellcolor{gray!40}\bfseries 57.0\% / 105k \\
\midrule
\multicolumn{7}{l}{\textbf{gemini-3-flash}} \\
\midrule
BigNum     &  62 &  6.5\% / 24k &  6.5\% / 35k &  8.1\% / 49k &  8.1\% / 63k &  9.7\% / 78k \\
VerifCogen & 172 & 25.6\% / 11k & 33.1\% / 17k & 39.0\% / 22k & 43.6\% / 28k & 45.9\% / 33k \\
Verina     & 189 & 24.9\% / 15k & 30.7\% / 23k & 38.1\% / 32k & 39.2\% / 42k & 40.2\% / 52k \\
\cellcolor{gray!10}\textbf{Combined} &
\cellcolor{gray!10}\bfseries 423 &
\cellcolor{gray!10}\bfseries 22.5\% / 15k &
\cellcolor{gray!10}\bfseries 28.1\% / 22k &
\cellcolor{gray!10}\bfseries 34.0\% / 30k &
\cellcolor{gray!10}\bfseries 36.4\% / 39k &
\cellcolor{gray!40}\bfseries 38.1\% / 48k \\
\midrule
\multicolumn{7}{l}{\textbf{gemini-3.1-flash-lite}} \\
\midrule
BigNum     &  62 &  3.2\% /  7k &  3.2\% / 11k &  3.2\% / 15k &  3.2\% / 19k &  3.2\% / 23k \\
VerifCogen & 172 & 21.5\% /  6k & 25.0\% / 11k & 26.7\% / 16k & 27.9\% / 21k & 31.4\% / 26k \\
Verina     & 189 & 11.1\% /  7k & 21.2\% / 12k & 24.9\% / 17k & 25.4\% / 22k & 25.4\% / 28k \\
\cellcolor{gray!10}\textbf{Combined} &
\cellcolor{gray!10}\bfseries 423 &
\cellcolor{gray!10}\bfseries 14.2\% /  6k &
\cellcolor{gray!10}\bfseries 20.1\% / 11k &
\cellcolor{gray!10}\bfseries 22.5\% / 16k &
\cellcolor{gray!10}\bfseries 23.2\% / 21k &
\cellcolor{gray!40}\bfseries 24.6\% / 26k \\
\bottomrule
\end{tabular}
}
\end{table}

\begin{table}
\caption[gpt-5.4 cumulative solves on \texttt{bignum} every five LLM calls.]{gpt-5.4 cumulative solves on \texttt{bignum}, sampled every five LLM calls through $K = 50$. Highlighted rows indicate the $K=20$ to $30$ range, where 22 of the eventual 54 solves occur.}
\label{tab:gpt-bignum-kcurve}
\renewcommand{\arraystretch}{1.15}
\centering
\begin{tabular}{r r r r}
\toprule
$\boldsymbol{K}$ & \textbf{solved / 62} & \textbf{\%} & \textbf{Change} \\
\midrule
 5 &  9 & 14.5\% & $+9$ \\
10 & 12 & 19.4\% & $+3$ \\
15 & 15 & 24.2\% & $+3$ \\
20 & 20 & 32.3\% & $+5$ \\
\rowcolor{gray!15} 25 & 31 & 50.0\% & $+11$ \\
\rowcolor{gray!15} 30 & 42 & 67.7\% & $+11$ \\
35 & 50 & 80.6\% & $+8$ \\
40 & 52 & 83.9\% & $+2$ \\
45 & 53 & 85.5\% & $+1$ \\
50 & 54 & 87.1\% & $+1$ \\
\bottomrule
\end{tabular}
\end{table}

\chapter{Search Methods for Lean Verification}
\label{ch:search}

\section{Searching Over Partial Proofs}
\label{sec:partial-proofs}

\subsection{Motivation}

The agent loop of Section~\ref{sec:agent-baseline} substantially improves over the Vericoder of Section~\ref{sec:vericoder}, and extended to $K = 50$ in Section~\ref{subsec:longer-context}, GPT-5.4 reaches 95.0\% on the combined 423-spec set and 87.1\% on \texttt{bignum}, effectively saturating the benchmark.
However, this saturation is confined to the strongest tested model and incurs significant token cost. At $K = 50$, GPT-5.4-mini reaches only 17.7\% on \texttt{bignum}, picking up just three additional solves on this subset between $K = 10$ and $K = 50$. Each \texttt{bignum} spec at $K = 50$ averages 139k unique tokens for GPT-5.4 and 207k unique tokens for GPT-5.4-mini.

A sign that faster solves are possible is the shape of GPT-5.4's \texttt{bignum} solve curve (Table~\ref{tab:gpt-bignum-kcurve}): the cumulative rate climbs slowly through $K = 20$ at 32.3\%, then jumps to 67.7\% across $K = 20 \to 30$, before flattening to its $K = 50$ ceiling of 87.1\%.
One possible reason for this shape is \emph{context-induced anchoring}: once an agent starts attempting a strategy in the first few turns, subsequent turns are more likely to extend and reinforce that strategy rather than reconsider it~\cite{10.48550/arxiv.2305.13534,10.48550/arxiv.2412.06593}. This may lead to an initial plateau before the model switches directions and constructs a solve using a more viable strategy. For smaller models, the agent may continue to attempt tactics within a fixed strategy while a more successful direction is never explored.

Tree-search methods are a general family of algorithms that maintain multiple candidate states, choosing at each step which one to extend. Applied to proofs, tree search has been a dominant approach across years of work, and naturally serves as a counter to such anchoring. In the standard setup, \emph{nodes} of the search tree are partial-proof states (a tactic state with its open goals, or one abstraction layer above such as a sketch or subgoal frontier), and the goal is to reach a node corresponding to a fully written, verified proof. A search algorithm then governs which node to expand next.

Previous proof-search systems mostly follow a score-then-expand pattern. An \emph{evaluation signal}, either a hand-coded heuristic or a learned value head, assigns each node a numerical score; a \emph{selection rule}, such as PUCT in MCTS-style systems like AlphaProof~\cite{Hubert_2025} or best-first-by-policy in BFS-Prover~\cite{10.48550/arxiv.2502.03438}, uses these scores to pick the next node to expand. The systems differ mainly on the \emph{granularity of expansion}, ranging from one Lean tactic per edge in MCTS to an entire proof sketch or subgoal in draft-then-fill approaches~\cite{10.48550/arxiv.2509.22819,10.48550/arxiv.2504.21801}.

We hypothesize that LLMs can offer a more flexible alternative: through direct prompting, we can ask a model to reason about high-level proof direction, as opposed to just attempting implementation. 
We aim to replace both the numerical evaluation signal and selection rule, with two primary model roles. A parent agent handles evaluation and selection, proposing the next high-level strategy to attempt. A subagent then handles expansion, attempting to extend the partial-proof state towards a complete, verified solution.

\subsection{Experimental Setup}

We design a \textit{state-based orchestrator}, which consists of a parent agent that directs the proof, and subagents which explore the proof-state tree. At the start, the parent uses an \texttt{explore\_variations} call to spawn multiple subagents for up to five turns each. Each subagent is primed with a different advice string drawn from a five-class strategy palette so that subagents pursue different proof directions. The palette deliberately includes recent automation tactics such as \texttt{grind}, which are powerful techniques for automatically closing goals but under-represented in older pretraining data.

Each subagent branches off the parent's conversation, inheriting the full history of prior decisions and the current partial-proof state, which starts at the original spec. Like the agent harness, subagents are equipped with \texttt{search\_mathlib} and \texttt{submit\_code} tools to query mathlib and attempt proofs. On a verified \texttt{submit\_code} the proof is returned and the parent exits; otherwise each subagent returns a debrief summarizing what it tried and what blocked it. A subagent that judges its direction dead can also \texttt{abandon} early, returning a debrief before using all turns.

Given the subagent debriefs, the parent can choose to update the base state against which replacements are generated. \texttt{update\_base} commits a partial-proof state surfaced by one of the subagents, and \texttt{undo\_base} reverts to the previous base. Each \texttt{update\_base} and \texttt{undo\_base} step walks an explicit proof-state tree implicitly created by the subagents: the new base may carry fewer \texttt{sorry} holes if the subagent closed subgoals, or more if the replacement introduces auxiliary lemmas with their own \texttt{sorry} placeholders. Subsequent subagents fill replacements against the new base rather than the original spec, so the parent's commits act as a decomposition mechanism: each committed step narrows or sharpens the remaining work for the next round. After optional base update or undo, \texttt{explore\_variations} is called again to dispatch a new round of subagents with refined advice. The parent is advised to include specific lemma names discovered from previous subagents when constructing new advice.

The parent loop repeats until a complete proof verifies, or a budget of 50 LLM calls is exhausted, accumulated across all parent and subagent turns. We run experiments with GPT-5.4 and GPT-5.4-mini, the two best-performing LLMs from the agent baseline, again at medium reasoning effort on the same 423-spec set used in Section~\ref{sec:agent-baseline}.

\subsection{Initial Results}
\label{subsec:orch-v1-initial}

At $K = 50$, the orchestrator solves 287 / 423 specs (67.8\%) with GPT-5.4 and 210 / 423 (49.6\%) with GPT-5.4-mini. Against the agent baseline at $K = 50$, the orchestrator trails by 27.2\% on GPT-5.4 and by 7.4\% on GPT-5.4-mini.

The parent loop's tool-call distribution confirms that the orchestrator runs the intended pipeline. Table~\ref{tab:orch-v1-parent-tools} gives the distribution of parent tool calls over the $K=50$ budget and the share of specs which use each tool at least once.
The parent is instructed (but not forced) to open with \texttt{explore\_variations}, and GPT-5.4 follows this instruction exactly, with all specs using \texttt{explore\_variations}. A minor instruction-following lapse is noticed for GPT-5.4-mini, which immediately calls \texttt{update\_base} for seven specs at first. One of these specs succeeds and exits, leading to the count of 422 specs using \texttt{explore\_variations}.

\begin{figure}
\centering
\includegraphics[width=0.95\textwidth]{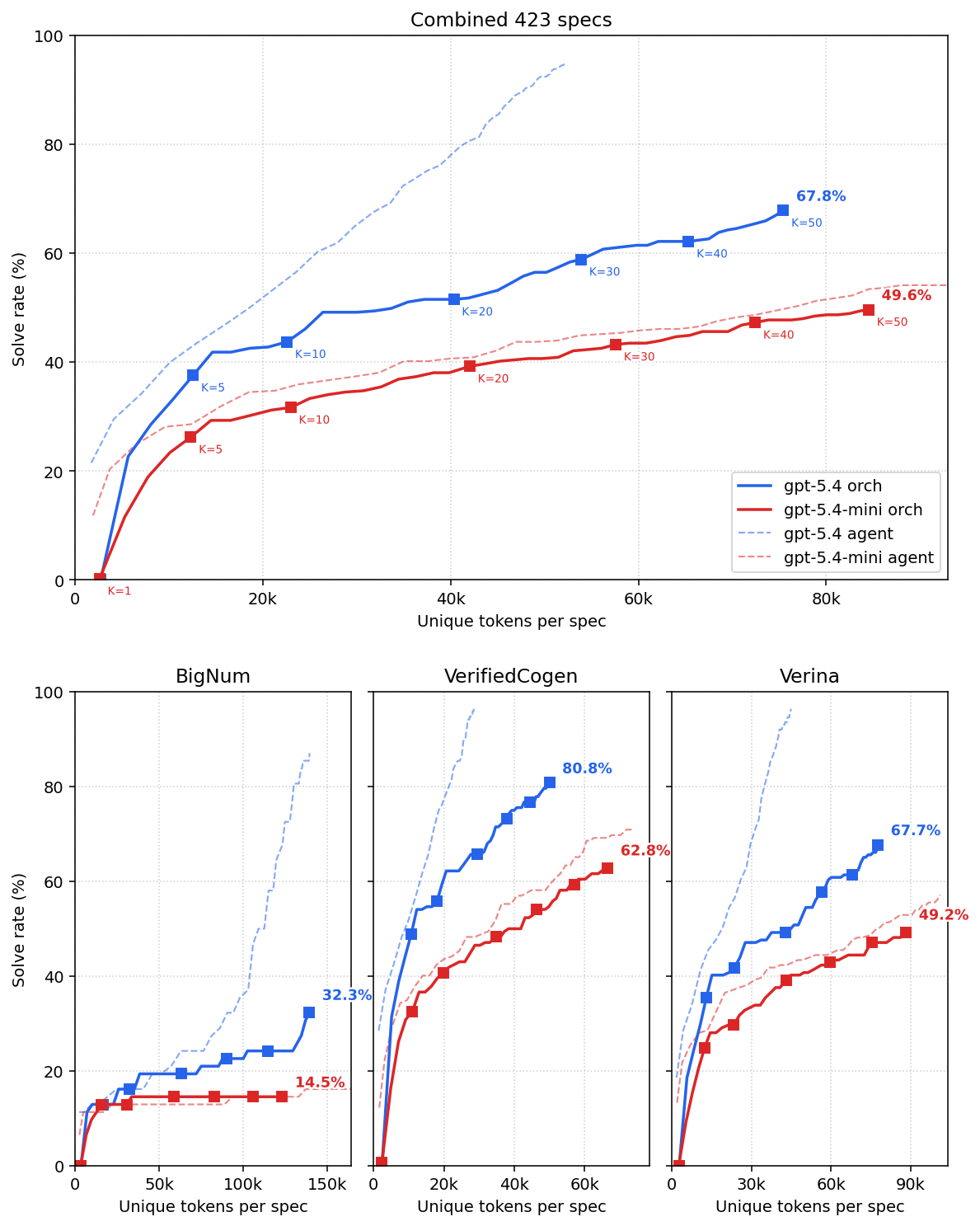}
\caption[State-based orchestrator solve rate vs.\ unique tokens, full set and per subset.]{State-based orchestrator solve rate against unique tokens, plotted alongside Section~\ref{sec:agent-baseline} agent baseline, for gpt-5.4 and gpt-5.4-mini. \emph{Top}: full 423-spec set. \emph{Bottom row, left to right}: \texttt{bignum}, \texttt{verified\_cogen}, and \texttt{verina} subsets. Orchestrator curves start later along the token axis because the first \texttt{explore\_variations} parent call is made before any subagent submits.}
\label{fig:orch-v1-pareto}
\end{figure}

Figure~\ref{fig:orch-v1-pareto} plots solve rate against unique tokens. The $K=1$ solve rate is close to zero, as the parent has an initial reasoning turn before dispatching subagents.
After the first turn, the solve rate shows its fastest increase through $K=6$, corresponding to the parent turn plus five subagent turns, trailing the agent by only 5.0\% for GPT-5.4 and 2.6\% for GPT-5.4-mini. Table~\ref{tab:orch-v1-kgrid} reports per-subset pass rates and per-spec tokens up to $K=50$.

A large fraction of solves occur in the first \texttt{explore\_variations}. For GPT-5.4, 52.2\% of specs finish in exactly one parent turn (a variation from the first \texttt{explore\_variations} submits a verified proof); for GPT-5.4-mini this share is 37.8\%; specs solved in the first \texttt{explore\_variations} make up at least 50\% of solved specs for each model. Within the first \texttt{explore\_variations}, most wins come from the first subagent out of up to four dispatched. The first subagent by itself generates a 41.8\% solve rate for GPT-5.4 and 29.1\% for GPT-5.4-mini.

\begin{table}
\caption[State-based orchestrator pass rate at varying $K$, by subset.]{State-based orchestrator pass rate at varying LLM-call budgets $K$, for gpt-5.4 and gpt-5.4-mini. Each cell shows pass rate~/ average unique tokens per spec at that $K$.}
\label{tab:orch-v1-kgrid}
\renewcommand{\arraystretch}{1.15}
\centering
\scalebox{0.85}{
\begin{tabular}{l r r r r r >{\columncolor{gray!10}\bfseries}r}
\toprule
\textbf{Subset} & \textbf{n} & \textbf{$K{=}10$} & \textbf{$K{=}20$} & \textbf{$K{=}30$} & \textbf{$K{=}40$} & \textbf{$K{=}50$} \\
\midrule
\multicolumn{7}{l}{\textbf{gpt-5.4}} \\
\midrule
BigNum     &  62 & 16.1\% / 32k & 19.4\% / 63k & 22.6\% / 90k  & 24.2\% / 115k & 32.3\% / 139k \\
VerifCogen & 172 & 55.8\% / 18k & 65.7\% / 30k & 73.3\% / 38k  & 76.7\% / 45k  & 80.8\% / 50k \\
Verina     & 189 & 41.8\% / 24k & 49.2\% / 43k & 57.7\% / 57k  & 61.4\% / 68k  & 67.7\% / 78k \\
\cellcolor{gray!10}\textbf{Combined} &
\cellcolor{gray!10}\bfseries 423 &
\cellcolor{gray!10}\bfseries 43.7\% / 23k &
\cellcolor{gray!10}\bfseries 51.5\% / 40k &
\cellcolor{gray!10}\bfseries 58.9\% / 54k &
\cellcolor{gray!10}\bfseries 62.2\% / 65k &
\cellcolor{gray!40}\bfseries 67.8\% / 75k \\
\midrule
\multicolumn{7}{l}{\textbf{gpt-5.4-mini}} \\
\midrule
BigNum     &  62 & 12.9\% / 31k & 14.5\% / 59k & 14.5\% / 82k  & 14.5\% / 106k & 14.5\% / 123k \\
VerifCogen & 172 & 40.7\% / 20k & 48.3\% / 35k & 54.1\% / 46k  & 59.3\% / 57k  & 62.8\% / 67k \\
Verina     & 189 & 29.6\% / 23k & 39.2\% / 43k & 42.9\% / 60k  & 47.1\% / 75k  & 49.2\% / 88k \\
\cellcolor{gray!10}\textbf{Combined} &
\cellcolor{gray!10}\bfseries 423 &
\cellcolor{gray!10}\bfseries 31.7\% / 23k &
\cellcolor{gray!10}\bfseries 39.2\% / 42k &
\cellcolor{gray!10}\bfseries 43.3\% / 58k &
\cellcolor{gray!10}\bfseries 47.3\% / 72k &
\cellcolor{gray!40}\bfseries 49.6\% / 85k \\
\bottomrule
\end{tabular}
}
\end{table}

\begin{table}
\caption[Parent tool-call distribution for state-based orchestrator.]{Parent-loop tool-call distribution over the $K=50$ budget across all 423 specs. \emph{Calls} counts total invocations; \emph{\% of calls} is the share within each model; \emph{Specs} counts unique specs on which the tool is called at least once.}
\label{tab:orch-v1-parent-tools}
\renewcommand{\arraystretch}{1.15}
\centering
\begin{tabular}{l r r r r}
\toprule
\textbf{Tool} & \textbf{Calls} & \textbf{\% of calls} & \textbf{Specs} & \textbf{\% of specs} \\
\midrule
\multicolumn{5}{l}{\textbf{gpt-5.4}} \\
\midrule
\texttt{explore\_variations} &  704 & 72.9\% & 423 & 100.0\% \\
\texttt{update\_base}        &  195 & 20.2\% & 152 &  35.9\% \\
\texttt{undo\_base}          &   67 &  6.9\% &  54 &  12.8\% \\
\midrule
\multicolumn{5}{l}{\textbf{gpt-5.4-mini}} \\
\midrule
\texttt{explore\_variations} & 1087 & 66.1\% & 422 &  99.8\% \\
\texttt{update\_base}        &  495 & 30.1\% & 223 &  52.7\% \\
\texttt{undo\_base}          &   62 &  3.8\% &  54 &  12.8\% \\
\bottomrule
\end{tabular}
\end{table}

\subsection{Subagent Search Behavior}
\label{subsec:orch-v1-subagent}

To understand how the orchestrator budget is spent, we analyze how subagents allocate their five-turn budget between \texttt{search\_mathlib} and \texttt{submit\_code} calls. Table~\ref{tab:orch-v1-subagent-tools} reports per-variation totals separately for the first and second \texttt{explore\_variations} call within each spec, alongside the corresponding agent baseline from Section~\ref{sec:agent-baseline}.

For GPT-5.4, the search-to-submit ratio inside a first-explore variation is 11.5, more than double the agent's ratio of 5.3 over a ten-turn budget. The ratio remains elevated in the second explore at 8.2, even though the parent has already received one round of debriefs and can pass refined advice. GPT-5.4-mini's first-explore ratio is 9.5 against an agent ratio of 7.2, dropping to 4.7 in the second explore.

These high search ratios, combined with the five-turn subagent budget, lead to low submission counts: a variation averages roughly one \texttt{submit\_code} call across all four cohorts (0.77 to 1.14). Each subagent inherits the parent's prompt and partial-proof state but no tool history, so its early turns are dominated by running lemma searches, overlapping with searches other variations have already performed in prior rounds. Whereas the agent in Section~\ref{sec:agent-baseline} can use early searches across all rounds, every orchestrator dispatch leads to fresh searches in subagents.

This motivates allowing subagent context to be preserved across parent rounds, rather than forcing each variation to rebuild it from scratch. We extend the parent's toolset with a \texttt{resume\_variations} call: instead of spawning fresh subagents, the parent picks one or more prior subagents and continues them, optionally with refreshed advice consolidated from all current debriefs. The parent retains discretion over which variations to resume, allowing it to prioritize promising subagents while conserving LLM-call budget by not resuming the rest.

The partial progress mechanism of \texttt{update\_base} and \texttt{undo\_base} remains unchanged. However, committing a new base implicitly invalidates any prior subagent context, since the partial-proof target a resumed subagent was working against no longer matches. After an \texttt{update\_base} or \texttt{undo\_base}, we make previous subagents ineligible for resume and require the parent to dispatch fresh subagents.

\begin{table}
\caption[State-based orchestrator subagent tool-call totals per variation.]{Average tool-call totals per subagent variation over its five-turn budget, split by first or second parent \texttt{explore\_variations} call. \emph{Turns/Var} is the average number of active turns per variation (out of five). \emph{Ratio} is total \texttt{search\_mathlib} divided by total \texttt{submit\_code} across the variation. Agent rows are ten-turn totals from the Section~\ref{sec:agent-baseline} baseline (Table~\ref{tab:tool-choice-summary}).}
\label{tab:orch-v1-subagent-tools}
\renewcommand{\arraystretch}{1.15}
\centering
\begin{tabular}{l l r r r r r}
\toprule
\textbf{Model} & \textbf{Block} & \textbf{\# Var} & \textbf{Turns/Var} & \textbf{Search} & \textbf{Submit} & \textbf{Ratio} \\
\midrule
\textbf{gpt-5.4}      & agent (10 turns) & --- & ---  & 17.53 & 3.48 &  5.29 \\
\textbf{gpt-5.4}      & 1st explore      & 982 & 4.22 & 11.24 & 0.98 & 11.47 \\
\textbf{gpt-5.4}      & 2nd explore      & 582 & 4.56 &  9.36 & 1.14 &  8.21 \\
\midrule
\textbf{gpt-5.4-mini} & agent (10 turns) & --- & ---  & 13.77 & 1.93 &  7.17 \\
\textbf{gpt-5.4-mini} & 1st explore      & 894 & 4.38 &  7.30 & 0.77 &  9.48 \\
\textbf{gpt-5.4-mini} & 2nd explore      & 511 & 4.20 &  4.25 & 0.91 &  4.67 \\
\bottomrule
\end{tabular}
\end{table}

\subsection{Final Results}

\begin{figure}
\centering
\includegraphics[width=0.95\textwidth]{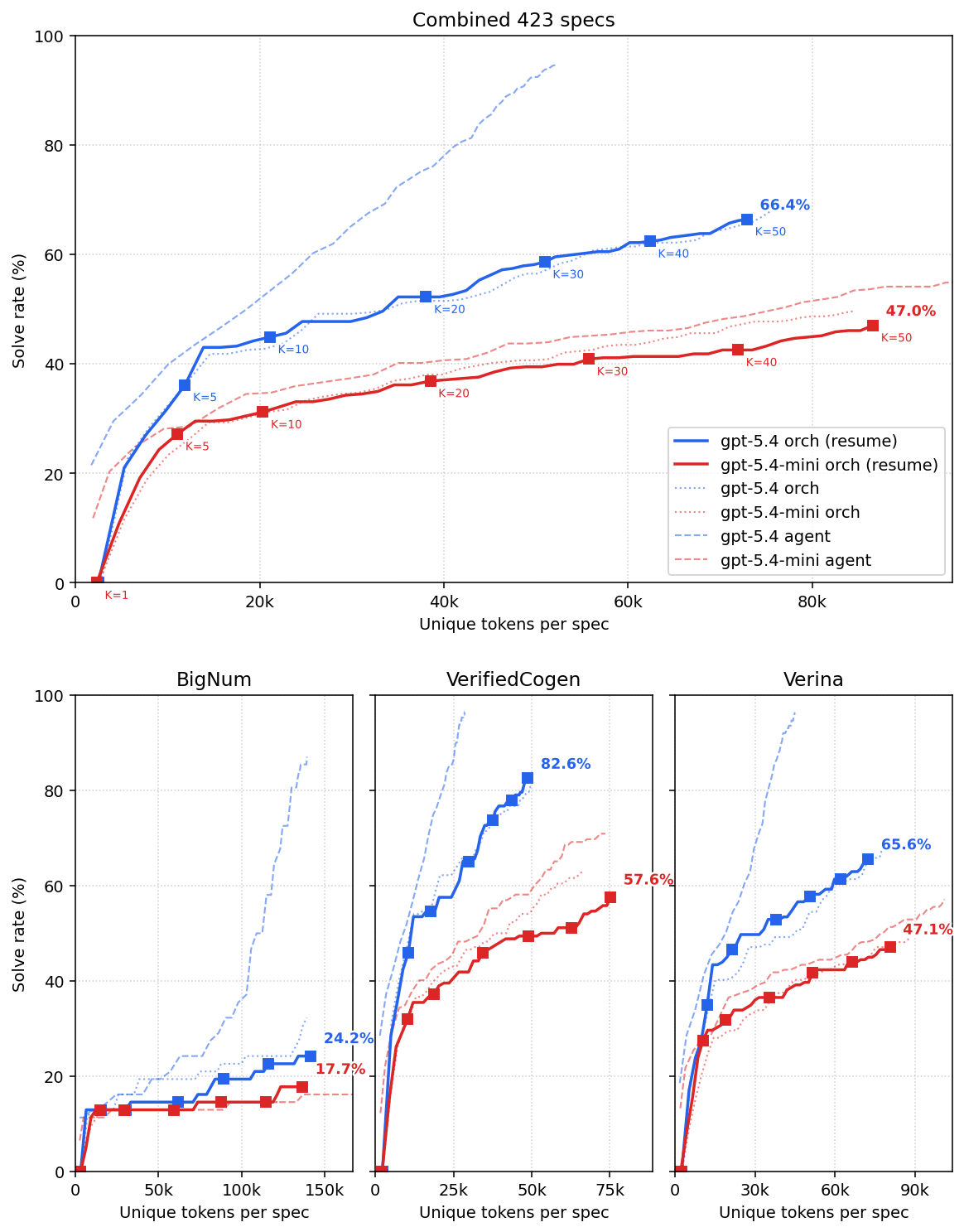}
\caption[State-based + resume orchestrator solve rate vs.\ unique tokens, full set and per subset.]{State-based + resume orchestrator solve rate against unique tokens, plotted alongside Section~\ref{subsec:orch-v1-initial} no-resume predecessor and Section~\ref{sec:agent-baseline} agent baseline, for gpt-5.4 and gpt-5.4-mini. \emph{Top}: full 423-spec set. \emph{Bottom row, left to right}: \texttt{bignum}, \texttt{verified\_cogen}, and \texttt{verina} subsets.}
\label{fig:orch-v1-resume-pareto}
\end{figure}

Adding \texttt{resume\_variations} to the parent toolset negatively affects the combined solve rate.
At $K=50$, GPT-5.4 drops 1.4\% to land at 66.4\% (281 / 423), while GPT-5.4-mini regresses 2.6\% to 47.0\% (199 / 423). Figure~\ref{fig:orch-v1-resume-pareto} plots the resume-enabled orchestrator against the Section~\ref{subsec:orch-v1-initial} no-resume orchestrator and the Section~\ref{sec:agent-baseline} agent, showing small gains and losses between models and subsets. Table~\ref{tab:orch-v1-resume-kgrid} reports per-subset pass rates and per-spec tokens up to $K=50$.

Table~\ref{tab:orch-v1-resume-parent-tools} reports the new parent tool-call distribution. GPT-5.4-mini makes heavier use of \texttt{resume\_variations}, accounting for 21.7\% of parent calls compared to 11.0\% for GPT-5.4.
\texttt{explore\_variations}, \texttt{update\_base}, and \texttt{undo\_base} each drop relative to the no-resume distribution in Table~\ref{tab:orch-v1-parent-tools}.
This difference in usage rate also translates into per-spec coverage: GPT-5.4-mini uses resume on 53.9\% of specs, while GPT-5.4 resumes on only 21.3\%.

The intended design of \texttt{resume\_variations} was for subagents granted more turns to shift from searching to submitting.
Table~\ref{tab:orch-v1-resume-subagent-tools} shows the opposite: for the second parent turn when resume is chosen, \texttt{search\_mathlib} accounts for 91.9\% of subagent calls on GPT-5.4 and 93.0\% on GPT-5.4-mini, higher than both the corresponding fresh second-explore rates and the first-explore rates. A possible reason is that the advice provided to resumed subagents contains methods and lemma names across the first round of subagents, leading resumed subagents to search again for mentioned lemmas instead of refining submissions.

\begin{table}
\caption[State-based + resume orchestrator pass rate at varying $K$, by subset.]{State-based + resume orchestrator pass rate at varying LLM-call budgets $K$, for gpt-5.4 and gpt-5.4-mini. Each cell shows pass rate~/ average unique tokens per spec at that $K$.}
\label{tab:orch-v1-resume-kgrid}
\renewcommand{\arraystretch}{1.15}
\centering
\scalebox{0.85}{
\begin{tabular}{l r r r r r >{\columncolor{gray!10}\bfseries}r}
\toprule
\textbf{Subset} & \textbf{n} & \textbf{$K{=}10$} & \textbf{$K{=}20$} & \textbf{$K{=}30$} & \textbf{$K{=}40$} & \textbf{$K{=}50$} \\
\midrule
\multicolumn{7}{l}{\textbf{gpt-5.4}} \\
\midrule
BigNum     &  62 & 12.9\% / 30k & 14.5\% / 62k & 19.4\% / 89k  & 22.6\% / 116k & 24.2\% / 142k \\
VerifCogen & 172 & 54.7\% / 18k & 65.1\% / 30k & 73.8\% / 38k  & 77.9\% / 44k  & 82.6\% / 49k \\
Verina     & 189 & 46.6\% / 21k & 52.9\% / 38k & 57.7\% / 51k  & 61.4\% / 62k  & 65.6\% / 72k \\
\cellcolor{gray!10}\textbf{Combined} &
\cellcolor{gray!10}\bfseries 423 &
\cellcolor{gray!10}\bfseries 44.9\% / 21k &
\cellcolor{gray!10}\bfseries 52.2\% / 38k &
\cellcolor{gray!10}\bfseries 58.6\% / 51k &
\cellcolor{gray!10}\bfseries 62.4\% / 62k &
\cellcolor{gray!40}\bfseries 66.4\% / 73k \\
\midrule
\multicolumn{7}{l}{\textbf{gpt-5.4-mini}} \\
\midrule
BigNum     &  62 & 12.9\% / 29k & 12.9\% / 59k & 14.5\% / 88k  & 14.5\% / 114k & 17.7\% / 137k \\
VerifCogen & 172 & 37.2\% / 19k & 45.9\% / 34k & 49.4\% / 49k  & 51.2\% / 63k  & 57.6\% / 75k \\
Verina     & 189 & 31.7\% / 19k & 36.5\% / 35k & 41.8\% / 52k  & 43.9\% / 67k  & 47.1\% / 81k \\
\cellcolor{gray!10}\textbf{Combined} &
\cellcolor{gray!10}\bfseries 423 &
\cellcolor{gray!10}\bfseries 31.2\% / 20k &
\cellcolor{gray!10}\bfseries 36.9\% / 39k &
\cellcolor{gray!10}\bfseries 40.9\% / 56k &
\cellcolor{gray!10}\bfseries 42.6\% / 72k &
\cellcolor{gray!40}\bfseries 47.0\% / 87k \\
\bottomrule
\end{tabular}
}
\end{table}

\begin{table}
\caption[Parent tool-call distribution for state-based + resume orchestrator.]{Parent-loop tool-call distribution over the $K=50$ budget across all 423 specs, with \texttt{resume\_variations} enabled. \emph{Calls} counts total invocations; \emph{\% of calls} is the share within each model; \emph{Specs} counts unique specs on which the tool is called at least once.}
\label{tab:orch-v1-resume-parent-tools}
\renewcommand{\arraystretch}{1.15}
\centering
\begin{tabular}{l r r r r}
\toprule
\textbf{Tool} & \textbf{Calls} & \textbf{\% of calls} & \textbf{Specs} & \textbf{\% of specs} \\
\midrule
\multicolumn{5}{l}{\textbf{gpt-5.4}} \\
\midrule
\texttt{explore\_variations} & 617 & 66.6\% & 423 & 100.0\% \\
\texttt{update\_base}        & 160 & 17.3\% & 132 &  31.2\% \\
\texttt{resume\_variations}  & 102 & 11.0\% &  90 &  21.3\% \\
\texttt{undo\_base}          &  47 &  5.1\% &  40 &   9.5\% \\
\midrule
\multicolumn{5}{l}{\textbf{gpt-5.4-mini}} \\
\midrule
\texttt{explore\_variations} & 805 & 52.5\% & 423 & 100.0\% \\
\texttt{update\_base}        & 354 & 23.1\% & 187 &  44.2\% \\
\texttt{resume\_variations}  & 332 & 21.7\% & 228 &  53.9\% \\
\texttt{undo\_base}          &  42 &  2.7\% &  38 &   9.0\% \\
\bottomrule
\end{tabular}
\end{table}

\begin{table}
\caption[State-based + resume orchestrator subagent tool-call totals per variation.]{Average tool-call totals per subagent variation, split by parent call: first \texttt{explore\_variations}, second \texttt{explore\_variations}, or \texttt{resume\_variations} chosen as the second decision. \emph{Turns/Var} is the average number of active turns per variation; explore variations have a five-turn budget, while resume variations are cumulative with their source explore (up to ten turns total). \emph{Ratio} is total \texttt{search\_mathlib} divided by total \texttt{submit\_code} across the variation.}
\label{tab:orch-v1-resume-subagent-tools}
\renewcommand{\arraystretch}{1.15}
\centering
\begin{tabular}{l l r r r r r}
\toprule
\textbf{Model} & \textbf{Block} & \textbf{\# Var} & \textbf{Turns/Var} & \textbf{Search} & \textbf{Submit} & \textbf{Ratio} \\
\midrule
\textbf{gpt-5.4}      & 1st explore         & 1{,}001 & 4.27 &  9.31 & 0.89 & 10.49 \\
\textbf{gpt-5.4}      & 2nd explore         &     395 & 4.32 &  7.32 & 0.91 &  8.03 \\
\textbf{gpt-5.4}      & resume (as 2nd)     &     148 & 8.45 & 17.37 & 1.53 & 11.38 \\
\midrule
\textbf{gpt-5.4-mini} & 1st explore         &     950 & 4.40 &  6.11 & 0.65 &  9.35 \\
\textbf{gpt-5.4-mini} & 2nd explore         &     380 & 4.12 &  3.29 & 0.79 &  4.16 \\
\textbf{gpt-5.4-mini} & resume (as 2nd)     &     272 & 8.57 & 11.62 & 0.87 & 13.34 \\
\bottomrule
\end{tabular}
\end{table}

We ran small-batch evaluations on several additional orchestrator modifications:
\begin{itemize}
    \item \textbf{Dynamic turn budgets}: allocates varying turn budgets depending on the number of subagents, to keep each \texttt{explore\_variations} call count a similar amount towards the total budget
    \item \textbf{Skeptical LLM reviewer}: a second model reviews the debriefs from the subagents, returning a second proof feasibility signal to the parent agent
    \item \textbf{Submission Elo ranking}: submissions from subagents are ranked by an LLM judge and assigned Elo scores, with the top submissions across rounds exposed to the parent agent
\end{itemize}
On the spec batches we evaluated, none of these shifted the headline solve rate beyond run-to-run variance; we leave further exploration of these methods to future work.

\section{Searching Over Agent Contexts}
\label{sec:agent-contexts}

\subsection{Motivation}

In Section~\ref{sec:agent-baseline} we found that models perform well with longer context, allowing for more turns to search lemmas and refine proof attempts. The analysis in Section~\ref{subsec:agent-tools} and Section~\ref{subsec:orch-v1-subagent} breaks down the behavior of subagents across turns: early searches of lemmas in context help the model craft more submissions, and without this context subagents return to searching instead of submitting.
Compared to the agent loop's search-to-submit ratio of 5.3 for GPT-5.4 over the first ten turns, the ratio for empty-context subagents more than doubles to 11.5, despite half the turn budget.

The design of the state-based orchestrator takes nodes to be partial-proof states against which the next round of subagents is dispatched.
However, subagent contexts are not preserved between rounds; this throws away exactly the chain of mathlib queries, verifier diagnostics, and abandoned tactics that produced the updated state.

The reframing is to save the path (context), instead of just the node (partial proof).
The challenge is reasoning, not formatting: completing the proof needs the same evidence base that the prior subagent used.
Narrowing the proof simplifies the replacements to generate, but generating valid output is not the bottleneck, and the previous context implicitly carries the partial-proof progress.
The same logic applies for the parent, which can make better decisions looking back at the entire subagent context rather than per-round debriefs, which compress multiple subagent turns into a single paragraph.

Together, these observations motivate a design which gives first-class support to continuous context while preserving a tree-search structure.
We keep the parent-director / subagent-prover framework from Section~\ref{sec:partial-proofs}, including advice-primed subagents and parent-side access to the proof-state tree.
We change what an edge of the search tree carries: rather than only the partial-proof state at its endpoint, the edge inherits the parent's full message history plus the subagent's per-turn transcript.
Searching through multiple proof states creates a tree of branching contexts, each carrying reasoning history and an evidence base for further searches. 

\subsection{Experimental Setup}

We design a \textit{context-based orchestrator} with emphasis on context preservation. The parent agent starts by dispatching a single subagent, running for up to ten turns instead of five. Subagents succeed, fail, or abandon as usual and return a debrief, but the parent operates on top of this context rather than only the compressed debrief. The subagent transcript, including all tool calls, is recorded as an \textit{endpoint} on which the next parent decision can branch.

The parent is given two tools that operate on the available endpoints. \texttt{dispatch\_subagent} spawns a new subagent with a fresh advice block from the parent. \texttt{resume\_endpoint} instead re-enters a prior subagent's exact context, allowing it to continue with a fresh turn allowance. These two tools naturally build a tree: one creates a new branch, and the other continues an existing one. The parent may also choose to abandon a spec early if determined infeasible.

The total budget across parent and subagent turns remains at 50 LLM calls, and we again use GPT-5.4 and GPT-5.4-mini.

\subsection{Initial Results}

\begin{figure}
\centering
\includegraphics[width=0.95\textwidth]{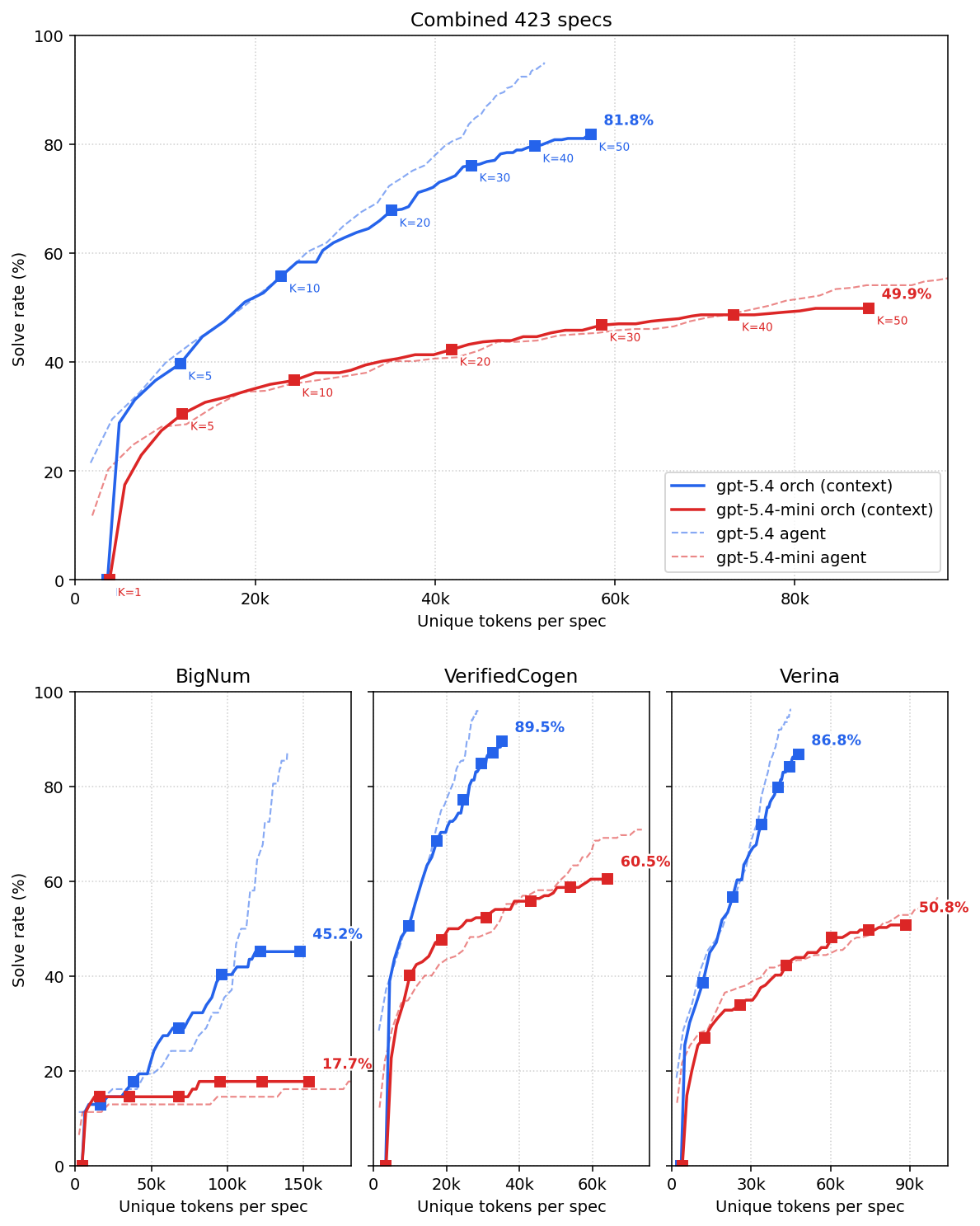}
\caption[Context-based orchestrator solve rate vs.\ unique tokens, full set and per subset.]{Context-based orchestrator solve rate against unique tokens, plotted alongside Section~\ref{sec:agent-baseline} agent baseline, for gpt-5.4 and gpt-5.4-mini. \emph{Top}: full 423-spec set. \emph{Bottom row, left to right}: \texttt{bignum}, \texttt{verified\_cogen}, and \texttt{verina} subsets.}
\label{fig:orch-v2-pareto}
\end{figure}

Our context-based orchestrator outperforms the state-based orchestrator from Section~\ref{subsec:orch-v1-initial} on GPT-5.4, reaching 81.8\% at $K=50$ (+14.0\%); on GPT-5.4-mini it remains nearly unchanged at 49.9\% (+0.3\%).
Figure~\ref{fig:orch-v2-pareto} plots solve rate against average unique tokens per spec, and Table~\ref{tab:orch-v2-kgrid} reports per-subset solve rates across $K$.

The context-based orchestrator initially produces solves at a similar rate as the previous one, as the first few turns correspond to the first subagent in both designs. The slope of the solve curve improves, matching the agent through ten LLM calls as opposed to quickly plateauing. This is not surprising as the new orchestrator acts as an agentic loop in its first subagent dispatch.
This matching remains through $K=10$ even with an initial parent turn, reaffirming the strength of initial reasoning.

However, the Pareto picture is less favorable at later iterations, with the GPT-5.4 orchestrator curve crossing below the agent curve after $K=10$ for the two easier subsets.
On the challenging \texttt{bignum} subset, the orchestrator maintains a slight lead over the agent through $K=20$, before falling off afterwards.

\begin{table}
\caption[Context-based orchestrator pass rate at varying $K$, by subset.]{Context-based orchestrator pass rate at varying LLM-call budgets $K$, for gpt-5.4 and gpt-5.4-mini. Each cell shows pass rate~/ average unique tokens per spec at that $K$.}
\label{tab:orch-v2-kgrid}
\renewcommand{\arraystretch}{1.15}
\centering
\scalebox{0.85}{
\begin{tabular}{l r r r r r >{\columncolor{gray!10}\bfseries}r}
\toprule
\textbf{Subset} & \textbf{n} & \textbf{$K{=}10$} & \textbf{$K{=}20$} & \textbf{$K{=}30$} & \textbf{$K{=}40$} & \textbf{$K{=}50$} \\
\midrule
\multicolumn{7}{l}{\textbf{gpt-5.4}} \\
\midrule
BigNum     &  62 & 17.7\% / 38k & 29.0\% / 68k & 40.3\% / 96k & 45.2\% / 122k & 45.2\% / 148k \\
VerifCogen & 172 & 68.6\% / 17k & 77.3\% / 25k & 84.9\% / 30k & 87.2\% / 33k  & 89.5\% / 35k \\
Verina     & 189 & 56.6\% / 23k & 72.0\% / 34k & 79.9\% / 40k & 84.1\% / 45k  & 86.8\% / 48k \\
\cellcolor{gray!10}\textbf{Combined} &
\cellcolor{gray!10}\bfseries 423 &
\cellcolor{gray!10}\bfseries 55.8\% / 23k &
\cellcolor{gray!10}\bfseries 67.8\% / 35k &
\cellcolor{gray!10}\bfseries 76.1\% / 44k &
\cellcolor{gray!10}\bfseries 79.7\% / 51k &
\cellcolor{gray!40}\bfseries 81.8\% / 57k \\
\midrule
\multicolumn{7}{l}{\textbf{gpt-5.4-mini}} \\
\midrule
BigNum     &  62 & 14.5\% / 36k & 14.5\% / 68k & 17.7\% / 95k & 17.7\% / 123k & 17.7\% / 154k \\
VerifCogen & 172 & 47.7\% / 19k & 52.3\% / 31k & 55.8\% / 43k & 58.7\% / 54k  & 60.5\% / 64k \\
Verina     & 189 & 33.9\% / 26k & 42.3\% / 43k & 48.1\% / 61k & 49.7\% / 75k  & 50.8\% / 89k \\
\cellcolor{gray!10}\textbf{Combined} &
\cellcolor{gray!10}\bfseries 423 &
\cellcolor{gray!10}\bfseries 36.6\% / 24k &
\cellcolor{gray!10}\bfseries 42.3\% / 42k &
\cellcolor{gray!10}\bfseries 46.8\% / 59k &
\cellcolor{gray!10}\bfseries 48.7\% / 73k &
\cellcolor{gray!40}\bfseries 49.9\% / 88k \\
\bottomrule
\end{tabular}
}
\end{table}

\subsection{Empty-Context Resume Behavior}

Within the orchestrator curves, there is a notable drop-off in solve rate after $K=10$.
As subagents run for ten turns, this drop-off point corresponds to the first parent decision, between resuming the first subagent and dispatching a new one.

Dispatching a new subagent currently starts a new context to avoid possible strategy lock-in. However, previously in Section~\ref{sec:partial-proofs} we found that short-context subagents search more heavily, even when given direct advice and lemma names from the parent agent, which negatively impacted performance of the previous orchestrator.

We analyze this by comparing the behavior of dispatched and resumed subagents after the first parent decision. We group parent decisions, and analyze the search and submit rates of resulting subagents. We find that resumed subagents succeed more often and average faster termination, but this is expected as the parent is more likely to resume subagents closer to a proof.
More notably, there is a significant difference in the ratio between searches and submits, with dispatched subagents again searching more than five times as often as resumed ones.
We infer that the first context of ten turns is likely to be useful base information: initial reasoning, lemma searches, and reusable early submissions. Removing the base context requires the model to rederive this from scratch, even when given context of what to try next.

The fix is to also have subagents branch off an explicit context endpoint. The parent may give a new direction synthesized from multiple endpoints, but previous information leading up to the endpoint is retained and available to build on. Previously, new subagents branched from the root, starting with only the parent advice. This change also allows more branching of contexts, allowing a tree of endpoints to take shape rather than independent lines of search.
The parent is also explicitly advised in the prompt to resume from endpoints rather than start new branches from the root, though the option remains available.

\begin{table}
\caption[Context-based orchestrator round-2 dispatch vs.\ resume subagent behavior.]{Behavioral comparison of round-2 dispatch and resume subagents (the parent's decision turn after a failed initial dispatch) on the context-based orchestrator with gpt-5.4. The dispatch column reports \texttt{dispatch\_subagent} calls, which run with an empty context. The resume column reports \texttt{resume\_endpoint} calls, which run off the round-1 context.}
\label{tab:orch-v2-r2-behavior}
\renewcommand{\arraystretch}{1.15}
\centering
\begin{tabular}{l r r}
\toprule
\textbf{Metric} & \textbf{Dispatch} & \textbf{Resume} \\
\midrule
n (round-2 cases)                              &     95 &     62 \\
Round-2 solve rate                             & 17.9\% & 38.7\% \\
\midrule
Mean turns to terminate                    &    7.8 &    4.1 \\
Mean \texttt{search\_mathlib} calls        &   15.1 &    2.3 \\
Mean \texttt{submit\_code} calls           &    1.9 &    2.7 \\
Mean \texttt{search\_mathlib} calls per turn   &    1.95 &   0.55 \\
Mean \texttt{submit\_code} calls per turn      &    0.24 &   0.64 \\
Search-to-submit ratio                         &    8.1 &    0.8 \\
\bottomrule
\end{tabular}
\end{table}

\subsection{Final Results}

\begin{figure}
\centering
\includegraphics[width=0.95\textwidth]{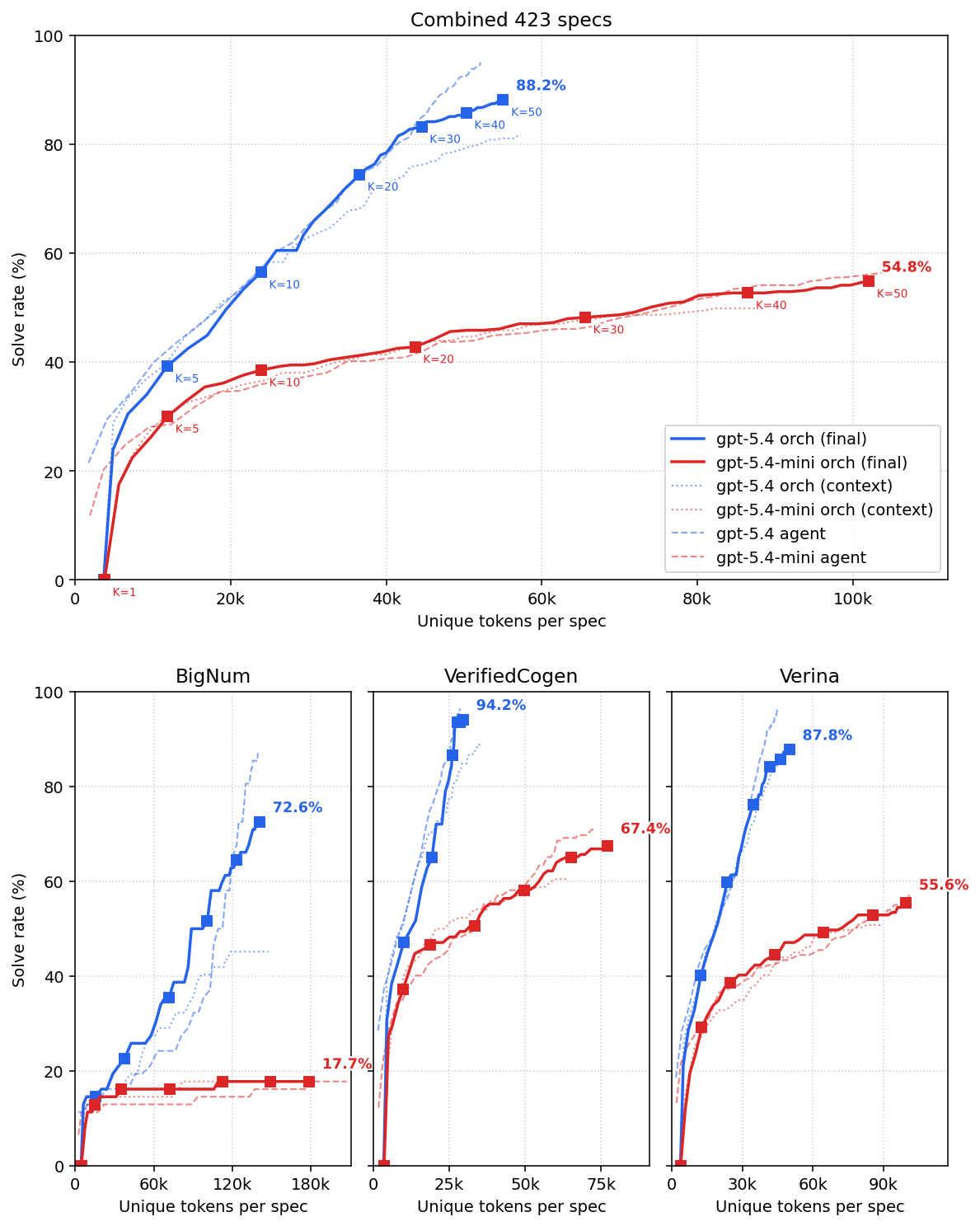}
\caption[Final orchestrator solve rate vs.\ unique tokens, full set and per subset.]{Final orchestrator solve rate against unique tokens, plotted alongside Section~\ref{sec:agent-contexts} context-based predecessor and Section~\ref{sec:agent-baseline} agent baseline, for gpt-5.4 and gpt-5.4-mini. \emph{Top}: full 423-spec set. \emph{Bottom row, left to right}: \texttt{bignum}, \texttt{verified\_cogen}, and \texttt{verina} subsets.}
\label{fig:orch-v3-pareto}
\end{figure}

The final context-based orchestrator reaches 88.2\% with GPT-5.4 and 54.8\% with GPT-5.4-mini at $K=50$, a $+6.4$\% and $+4.9$\% lift over the previous design on the same grid. Figure~\ref{fig:orch-v3-pareto} plots solve rate against average unique tokens per spec, and Table~\ref{tab:orch-v3-kgrid} reports per-subset solve rates across $K$.

Table~\ref{tab:orch-v3-r2-behavior} repeats the round-2 behavioral comparison from Table~\ref{tab:orch-v2-r2-behavior} for the final orchestrator.
The rate at which the parent chooses between dispatch and resume remains roughly similar.
The resumed subagent's success rate moderately increases, while dispatched subagents maintain a similar success rate as before.

More notably, the search-to-submit ratio for dispatch drops over 50\%, indicating a substantial shift in behavior: rather than searching mathlib from scratch, the subagent spends more of its turns on submissions because the source endpoint's prior search results are already in context.
The ratio remains higher than resume's 0.6, which is expected, as dispatch asks the subagent to go in a different direction and a fraction of its turns are spent reorienting.
Mean turns to termination also drops from 7.8 to 5.7, which improves the Pareto curve because successful specs no longer contribute tokens at higher rounds.

Our final orchestrator design is competitive with the agent baseline through $K=20$ on the combined 423-spec set, but solve rate slows past that point (Figure~\ref{fig:orch-v3-pareto}). The agent extended to $K=50$ reaches a higher solve rate on every GPT-5.4 subset, while for GPT-5.4-mini, the orchestrator and agent remain within a couple of percentage points.

The pattern varies in shape across subsets and models. On \texttt{verified\_cogen}, the easiest subset, the GPT-5.4 orchestrator incurs extra tokens in the first few turns and solves specs slower than the agent, only closing the gap when they both plateau.
On \texttt{verina}, the orchestrator stays close to the agent baseline at low $K$ but shows a drop-off in solve rate at higher $K$.
On the hardest benchmark, \texttt{bignum}, the orchestrator instead leads the agent through a significant portion of the token range, with the two curves crossing near 120k unique tokens per spec. At $K=50$, the GPT-5.4 agent reaches 87.1\% on \texttt{bignum} against the orchestrator's 72.6\%, roughly fifteen percentage points ahead.
For GPT-5.4-mini, the orchestrator shows small gains over the agent on some subsets and falls slightly behind on others, staying within a couple of percentage points.

We attribute this behavior to a tradeoff in intermediate parent reasoning.
On specs that require a strategy search across distinct proof approaches, the parent's ability to dispatch a new subagent with a different palette class lets it find a working approach faster than the agent's single-context exploration.
However, this hurts the orchestrator on specs that require deep iteration within a single approach, as every parent decision is a potential context switch that interrupts that iteration, and the agent's unbroken trajectory through $K=50$ converts more of these deep-iteration specs.

\begin{table}
\caption[Final orchestrator pass rate at varying $K$, by subset.]{Final orchestrator pass rate at varying LLM-call budgets $K$, for gpt-5.4 and gpt-5.4-mini. Each cell shows pass rate~/ average unique tokens per spec at that $K$.}
\label{tab:orch-v3-kgrid}
\renewcommand{\arraystretch}{1.15}
\centering
\scalebox{0.85}{
\begin{tabular}{l r r r r r >{\columncolor{gray!10}\bfseries}r}
\toprule
\textbf{Subset} & \textbf{n} & \textbf{$K{=}10$} & \textbf{$K{=}20$} & \textbf{$K{=}30$} & \textbf{$K{=}40$} & \textbf{$K{=}50$} \\
\midrule
\multicolumn{7}{l}{\textbf{gpt-5.4}} \\
\midrule
BigNum     &  62 & 22.6\% / 38k & 35.5\% / 72k & 51.6\% / 101k & 64.5\% / 123k & 72.6\% / 141k \\
VerifCogen & 172 & 65.1\% / 19k & 86.6\% / 26k & 93.6\% / 28k  & 93.6\% / 29k  & 94.2\% / 30k \\
Verina     & 189 & 59.8\% / 24k & 76.2\% / 35k & 84.1\% / 41k  & 85.7\% / 46k  & 87.8\% / 50k \\
\cellcolor{gray!10}\textbf{Combined} &
\cellcolor{gray!10}\bfseries 423 &
\cellcolor{gray!10}\bfseries 56.5\% / 24k &
\cellcolor{gray!10}\bfseries 74.5\% / 37k &
\cellcolor{gray!10}\bfseries 83.2\% / 45k &
\cellcolor{gray!10}\bfseries 85.8\% / 50k &
\cellcolor{gray!40}\bfseries 88.2\% / 55k \\
\midrule
\multicolumn{7}{l}{\textbf{gpt-5.4-mini}} \\
\midrule
BigNum     &  62 & 16.1\% / 35k & 16.1\% / 72k & 17.7\% / 113k & 17.7\% / 149k & 17.7\% / 179k \\
VerifCogen & 172 & 46.5\% / 19k & 50.6\% / 34k & 58.1\% / 50k  & 65.1\% / 65k  & 67.4\% / 77k \\
Verina     & 189 & 38.6\% / 25k & 44.4\% / 44k & 49.2\% / 65k  & 52.9\% / 85k  & 55.6\% / 100k \\
\cellcolor{gray!10}\textbf{Combined} &
\cellcolor{gray!10}\bfseries 423 &
\cellcolor{gray!10}\bfseries 38.5\% / 24k &
\cellcolor{gray!10}\bfseries 42.8\% / 44k &
\cellcolor{gray!10}\bfseries 48.2\% / 66k &
\cellcolor{gray!10}\bfseries 52.7\% / 86k &
\cellcolor{gray!40}\bfseries 54.8\% / 102k \\
\bottomrule
\end{tabular}
}
\end{table}

\begin{table}
\caption[Final orchestrator round-2 dispatch vs.\ resume subagent behavior.]{Behavioral comparison of round-2 dispatch and resume subagents (the parent's decision turn after a failed initial dispatch) on the final orchestrator with gpt-5.4. \texttt{dispatch\_subagent} typically branches from a non-root endpoint, so the dispatched subagent inherits the source endpoint's prior message history along with the parent's fresh advice.}
\label{tab:orch-v3-r2-behavior}
\renewcommand{\arraystretch}{1.15}
\centering
\begin{tabular}{l r r}
\toprule
\textbf{Metric} & \textbf{Dispatch} & \textbf{Resume} \\
\midrule
n (round-2 cases)                              &    100 &     70 \\
Round-2 solve rate                             & 18.0\% & 48.6\% \\
\midrule
Mean turns to terminate                        &    5.7 &    3.5 \\
Mean \texttt{search\_mathlib} calls            &    7.0 &    1.3 \\
Mean \texttt{submit\_code} calls               &    1.8 &    2.4 \\
Mean \texttt{search\_mathlib} calls per turn   &   1.23 &   0.39 \\
Mean \texttt{submit\_code} calls per turn      &   0.33 &   0.68 \\
Search-to-submit ratio                         &    3.8 &    0.6 \\
\bottomrule
\end{tabular}
\end{table}

\chapter{Conclusion}
\label{ch:conclusion}

We provide an updated evaluation for LLM-driven verified-code generation in Lean and devise search-based methods to improve verification performance.
Benchmarking recent frontier and open-weight models against the existing \texttt{vericoding-benchmark} Lean harness, closed-source US models show little year-over-year movement without reasoning, while open-weight models slightly edge up.
By replacing the original Vericoding harness with an agentic loop and incorporating mathlib search, solve rates significantly increase, scaling with additional LLM calls. GPT-5.4 hits 95.0\% on the 423-spec subset at $K=50$ LLM calls, nearly saturating the benchmark.
On top of the agent loop we build two orchestrators: a state-based design that picks which partial-proof state to extend next, and a context-based design that picks which complete subagent context to fork from.

We find that the benefit of search structure depends on the difficulty and width of the underlying problems.
A broad but shallow search, like the state-based orchestrator, quickly closes a wide swath of easier specs that are helped with an initial planning turn, but struggles to build up persistent progress on harder specs.
A context-aware tree search, like our final orchestrator design, is able to solve a range of more difficult problems that require strategies across distinct proof approaches, and does so faster than the agent baseline.
The tradeoff is that intermediate parent decisions interrupt the deep iteration that the agent baseline uses to crack the hardest specs, so the orchestrator closes those goals more slowly.

A second observation is that \texttt{vericoding-benchmark} is likely saturated by recent frontier models.
We evaluated on GPT-5.4 at medium reasoning, one generation behind the current flagship GPT-5.5, and not at xhigh, the maximum supported reasoning effort.
Contamination of the benchmark is also a possibility: while GPT-5.4's claimed knowledge cutoff of August 31, 2025 is before the release of \texttt{vericoding-benchmark} in September 2025, the model could have been exposed to the verification tasks in post-training pipelines.
Future models are increasingly likely to have the benchmark tasks incorporated into training, especially as interest in improving formal verification capabilities remains high.
More difficult formal-verification benchmarks, ideally with specs drawn from recent human-written code, are needed both to cleanly separate algorithmic progress from model capabilities, and to serve as a concrete test for those developing formal verification algorithms.

Real-world deployments of these systems also require different tradeoffs and considerations than the ones we make in this thesis.
We idealize provider pricing by counting unique input + output tokens, without always using optimal caching strategies (such as handling tool schema changes), and assume perfect prompt caching with free cache reads.
In practice, efficient harnesses may require more engineering and algorithmic considerations, and best practices for interfacing with model providers also change quickly.
A complementary direction to our custom orchestrator implementation is building search and dispatch on top of existing agent harnesses, inheriting their tool ecosystems and optimizations. However, this comes at the cost of less control over the underlying context, and being tied to nuances and development changes of an existing system.
In either case, the right design depends on the end user and the target application, not necessarily a single benchmark score.

Formal verification continues to grow as an area of research, thanks in part to improving model capabilities that lower the cost of writing and checking proofs.
In cybersecurity, mature verification tooling can harden existing codebases by formally ruling out vulnerabilities before the next exploit hits.
For developing reliable agents, machine-checkable correctness provides an automatic scaffold, lengthening the task horizon over which intermediate steps can be validated without a human in the loop.
At a longer horizon, automated formal verification could be used to verify AI systems themselves, complementing mechanistic interpretability by guaranteeing how a model behaves as opposed to explaining its internal logic.
Taken together, these problems make automatic formal verification a high-impact domain in the present and future, and one where continued algorithmic progress, alongside advances in tooling and testing, can deliver outsized returns.

\appendix
\defbibheading{bibintoc}{\chapter*{#1}\addcontentsline{toc}{backmatter}{\refname}} 
\printbibliography[title={\refname},heading=bibintoc]
\end{document}